\documentclass[10pt,journal,compsoc]{IEEEtran}

\ifCLASSOPTIONcompsoc
  \usepackage[nocompress]{cite}
\else
  \usepackage{cite}
\fi
\ifCLASSINFOpdf
\else
\fi
\usepackage{acro}				
\usepackage[utf8]{inputenc}		
\usepackage[font=footnotesize,labelfont=sf,textfont=sf]{caption}
\usepackage[font=footnotesize,labelfont=sf,textfont=sf]{subcaption}			
\usepackage{relsize}			
\usepackage{amsmath}			
\usepackage{multirow}			
\usepackage{mathtools}			
\usepackage[cal=boondoxo]{mathalfa}
\usepackage{amsfonts}
\usepackage{booktabs}
\usepackage{enumitem}
\usepackage{textcomp}

\DeclareAcronym{DT}{short = DT, long  = decision tree}
\DeclareAcronym{SS}{short = SS, long = soft search}
\DeclareAcronym{STP}{short = STP, long = soft training propagation}
\DeclareAcronym{SE}{short = SE, long = soft evaluation}
\DeclareAcronym{CT}{short = CT, long = computerized tomography}
\DeclareAcronym{EF}{short = EF, long = ejection fraction}
\DeclareAcronym{CMR}{short = CMR, long = cardiac magnetic resonance}
\DeclareAcronym{FN}{short = FN, long = false negatives}
\DeclareAcronym{FP}{short = FP, long = false positives}
\DeclareAcronym{TN}{short = TN, long = true negatives}
\DeclareAcronym{TP}{short = TP, long = true positives}
\DeclareAcronym{ID3}{short = ID3, long = Iterative Dichotomiser 3}
\DeclareAcronym{CDF}{short = CDF, long = cumulative density function}
\DeclareAcronym{ML}{short = ML, long = machine learning}
\DeclareAcronym{CV}{short = CV, long = cross-validation}
\DeclareAcronym{PLT}{short = PLT, long = piecewise-linear thresholds}
\DeclareAcronym{UDT}{short = UDT, long = uncertain decision tree}
\DeclareAcronym{JCGM}{long = Joint Committee for Guides in Metrology}

\newcommand{\specialcell}[2][c]{\begin{tabular}[#1]{@{}c@{}}#2\end{tabular}}

\begin{document}
%
\title{Decision Tree Learning for Uncertain \\ Clinical Measurements}
%
%
%
%

\author{Cecília Nunes, Hélène Langet, Mathieu De Craene, Oscar Camara, Bart Bijnens, Anders Jonsson%
\IEEEcompsocitemizethanks{\IEEEcompsocthanksitem C. Nunes, O. Camara, B. Bijnens and A. Jonsson are with the Department of Information and Communication Technologies, Universitat Pompeu Fabra, Barcelona, Spain. \protect \\
	E-mail: cecilia.nunes@upf.edu
	\IEEEcompsocthanksitem C. Nunes, H. Langet and M. De Craene are with Philips Research Medisys, Paris, France. \protect
	\IEEEcompsocthanksitem B. Bijnens is with ICREA, Barcelona, Spain.
	}
\thanks{\textcopyright  ~2018 IEEE.  Personal use of this material is permitted.  Permission from IEEE must be obtained for all other uses, in any current or future media, including reprinting/republishing this material for advertising or promotional purposes, creating new collective works, for resale or redistribution to servers or lists, or reuse of any copyrighted component of this work in other works. Manuscript received August 28, 2018.}
}

\markboth{(Preprint) IEEE Transactions on Knowledge and Data Engineering, Submitted for review, August~2019}%
{Nunes \MakeLowercase{\textit{et al.}}: Decision Tree Learning for Uncertain Clinical Measurements}
%



\IEEEtitleabstractindextext{%
\begin{abstract}

Clinical decision requires reasoning in the presence of imperfect data. \Acp{DT} are a well-known decision support tool, owing to their interpretability, fundamental in safety-critical contexts such as medical diagnosis. However, learning \acp{DT} from uncertain data leads to poor generalization, and generating predictions for uncertain data hinders prediction accuracy. Several methods have suggested the potential of probabilistic decisions at the internal nodes in making \acp{DT} robust to uncertainty. Some approaches only employ probabilistic thresholds during evaluation. Others also consider the uncertainty in the learning phase, at the expense of increased computational complexity or reduced interpretability. The existing methods have not clarified the merit of a probabilistic approach in the distinct phases of \ac{DT} learning, nor when the uncertainty is present in the training or the test data. We present a probabilistic \ac{DT} approach that models measurement uncertainty as a noise distribution, independently realized: (1)~when searching for the split thresholds, (2)~when splitting the training instances, and (3)~when generating predictions for unseen data. The soft training approaches (1, 2) achieved a regularizing effect, leading to significant reductions in \ac{DT} size, while maintaining accuracy, for increased noise. Soft evaluation (3) showed no benefit in handling noise.
\end{abstract}

}

\maketitle

\IEEEdisplaynontitleabstractindextext

%
\IEEEpeerreviewmaketitle

\IEEEraisesectionheading{\section{Introduction}\label{intro}}

\IEEEPARstart{C}{linical} data mining is the application of machine learning to medical databases as a tool to assist clinical research and decision. Its impact depends on one hand on the clinical pertinence of the models, but also on their predictive performance, interpretability, and appropriate integration in clinical software~\cite{Patel2009}. While many methodological successes exist, examples that affect patient management are seldom found~\cite{Clifton2015}. Measurement uncertainty limits the application of data mining, as it impairs model performance and generalizability~\cite{Roddick2003,Lavrac1998}. Improving data quality is however resource-intensive and often unfeasible~\cite{Karr2006}. We hypothesize that the acknowledgement of uncertainty, in particular by integrating domain-knowledge about the reliability of each measurement, can improve models and the leverage them as an asset for clinical research and decision making.

The uncertainty of a measurement reflects the lack of knowledge about its exact value, and is often caused by noise in the acquisition~\cite{jcgm2008evaluation}. Clinical measurement uncertainty originates from multiple sources including distinct diagnostic practices~\cite{Cios2002}, inter- and intra-observer variability~\cite{Singh2003}, manufacturer-dependent technology~\cite{Foley2012}, the use of distinct modalities for the same measurement~\cite{Lopez-Minguez2014}, or patient factors such as body habitus or claustrophobia, which affect the choice and the quality of an imaging test~\cite{genders2017quantitative}. E.g. in the determination of device size for left-atrial appendage closure, consistency between \ac{CT}, transesophageal echocardiography and angiography was observed in only 21.6\% of the cases~\cite{Lopez-Minguez2014}. In the estimation of \ac{EF}, the variability between \ac{CMR} and echocardiography resulted in 28\% of the population having opposing device eligibility~\cite{DeHaan2014}. Clinical reasoning involves making guideline-abiding decisions based on such unreliable data. In order to employ scientific evidence in practice, the experienced clinician assesses the reliability of each measurement, and integrates it with his/her training and experience. This endeavor is all the more challenging, considering the emphasis on internal validity of medical research, as opposed to external validity. The contrast between the scrutinous design of populations used for research and the actual populations where evidence is employed has been considered an obstacle to evidence-based medicine~\cite{green2014closing}.

\Acp{DT} are a knowledge-representation structure, where decisions at the internal nodes lead to a prediction at the leaves. They can be efficiently built using the well-known algorithms ID3~\cite{quinlan1979interactive}, C4.5~\cite{Quinlan1993}, CART~\cite{Breiman1984} or CHAID~\cite{kass1980exploratory}. Compared to other methods, \acp{DT} can be interpreted as a sequence of decisions. Although recent methods such as deep neural networks can offer better performance, their output is often a black box~\cite{Weng2017}. Interpretability is all the more necessary as recent European law secures the right to an explanation of all algorithmically-made decisions~\cite{EU}.

Learning \acp{DT} from noisy measurements can overfit to the noise and fail to generalize. Generating predictions for noisy instances can generate incorrect predictions. Each test at a \ac{DT} node compares a measurement with a \textit{hard threshold}, such that small errors can lead to the instance following an opposing path. Moreover, the distance between the measurement and the threshold is disregarded~\cite{Quinlan1987}. Several algorithms explore the idea of \textit{soft thresholds} to make \acp{DT} robust to the uncertainty~\cite{Quinlan1987,DvorakS07,Tsang2011,irsoy2012soft,Yuan1995}. Such approaches weight the contribution of all child branches to the prediction. Some methods focus on cognitive uncertainty, while others focus on statistical uncertainty or noise. The notion of \emph{fractional tuple} was first introduced in C4.5 for handling missing values~\cite{Quinlan1993}.

Fuzzy \acp{DT} use fuzzy logic to handle \textit{cognitive uncertainty}, which deals with inconsistencies in human reasoning~\cite{Yuan1995}. Using fuzzy \acp{DT} to handle statistical noise involves setting the fuzzy membership functions to express uncertainties around the \ac{DT} thresholds, known in advance. Alternatively, discretization algorithms can be used. But those methods are computationally costly~\cite{Wang2000}, and the accuracy of the resulting \acp{DT} strongly depends on the chosen algorithm~\cite{Segatori2017}.


Probabilistic \acp{DT}, on the other hand, focus on \textit{stochastic uncertainty} or random noise, arising from the unpredictable behavior of physical systems and measurement limitations:

\begin{itemize}[leftmargin=5mm]
	\item Quinlan~\cite{quinlan1990probabilistic} proposed a method using \acp{PLT} for evaluating a~\ac{DT}, assuming a two-modal uniform distribution for the noise. The parameters of this distribution are set using a statistical heuristic based on the training data. Dvor{\'{a}}k and Savick{\'{y}}~\cite{DvorakS07} employed a variation of this method, where the parameters were estimated through simulated annealing. Experiments in one dataset led to 2-3\% error rate reductions compared to CART~\cite{Breiman1984}, suggesting the potential of the method and the need for an evaluation on more data. No significant differences were found between the two parameter-estimation methods, but the authors highlight the computational cost of simulated annealing. The approach is applied to a finalized DT, so the uncertainty is not accounted for during training. 
	
	\item The \ac{UDT} algorithm~\cite{Tsang2011} extends the probabilistic splits to the training phase, assuming a Gaussian noise, and achieving accuracy gains in $10$ datasets. The method takes an oversampling strategy, where each measurement is replaced by $s$ points. The authors propose optimized searches to control the therefore increased training time. In \ac{UDT}, the same noise model is used for training and evaluation. In medical research and practice, however, the noise in the data used to obtain evidence can be very different from the noise in the data used in practice~\cite{green2014closing}. A learning algorithm ideally supports independent uncertainty models for the training data and the target data.
	
	\item Soft approaches have also been proposed for multivariate \acp{DT}. An algorithm~\cite{irsoy2012soft} employs logistic regression, treating the separation of data among the children as a classification problem. The method led to small accuracy gains in 10 classification datasets, with significant reductions in tree size. Hierarchical mixtures of experts are another multivariate tree architecture~\cite{jordan1994hierarchical}, where each test is a softmax linear combination of all variables, with performance gains compared to CART. Multivariate \acp{DT} can compactly model complex phenomena, and generally improve accuracy with fewer nodes. They are however intended to generate predictions that do not need to be understood, lacking the interpretability of univariate~\acp{DT}.	
\end{itemize}


The aforementioned approaches show the benefit of probabilistic \acp{DT} in improving \ac{DT} performance. However, the strategies used to model noise are either limited to the evaluation phase, or they do not use independent noise distributions for learning and evaluation. Moreover, the behavior of the methods upon increasing levels of noise was not investigated, and no distinction was made between the impact of noise in the training data and noise in the test data. A practical, flexible and well-understood approach for handling measurement uncertainty has not been established for~\ac{DT} learning.

This manuscript proposes a probabilistic \ac{DT} learning approach to handle uncertainty, modeled as a distribution of noise added to the real measurements. Unlike previous methods, the uncertainty distribution is orthogonally considered:
\begin{enumerate}[leftmargin=5mm]
	\item during training, when searching for the best threshold at each node, denoted \acf{SS};
	\item during training, in the propagation of the training data through the \ac{DT}, denoted \acf{STP};
	\item in the propagation of the test instances when evaluating a finalized tree, or \acf{SE}.
\end{enumerate}	
\noindent
To promote interpretability, we consider univariate \acp{DT} where each node contains a decision based on a single variable. The proposed \ac{SS} keeps the computational cost under control. We address the problem of integrating knowledge about the uncertainty coming from clinical experience and from the meta-analysis of clinical studies. As a proof-of-concept, we opted for a normal noise model, and evaluate its impact on the distinct learning phases. We also separately study the effect of corrupting the training or test data.

In the following, we introduce the ID3 and C4.5 algorithms and discuss a probabilistic interpretation. The manuscript then proceeds with a description of the proposed soft \ac{DT} algorithm components, followed by the experiments to illustrate and evaluate them.

	\section{Decision tree learning} \label{background}

Consider the input variable~$\mathbf{X}$ with dimensions $X_j$, $j = 1, ..., M$, and the output~$Y$, related by the unknown distribution~$P_{\mathbf{X}Y}(\mathbf{x}, y)$. Drawing samples from $P_{XY}$ composes the training dataset~$\mathcal{D}$. The supervised learning problem consists in learning a model from $\mathcal{D}$ that predicts $Y$ for an unseen sample~$\mathbf{x}$. \Acp{DT}~algorithms follow an algorithmic approach that does not attempt to learn~$P_{\mathbf{X}Y}(\mathbf{x}, y)$.

Learning an optimal \ac{DT} that maximizes generalization accuracy with a minimum number of decisions is NP-complete~\cite{Hyafil1976}. Although non-greedy methods exist for multivariate~\acp{DT}, locally-optimal approaches offer a good trade-off of accuracy and computational complexity. Notable examples include the \ac{ID3} \cite{Quinlan1986}, CART~\cite{Breiman1984} and C4.5~\cite{Quinlan1993}. The \ac{ID3} selects binary splits of numerical variables using the \textit{information gain}, and the C4.5 extends it to categorical features. For a survey on top-down \ac{DT} induction, refer to Rokach~\textit{et~al.}~\cite{Rokach2005}.  

\Acp{DT} are regularized by employing node \textit{pruning} algorithms. C4.5 employs a postpruning approach that cuts branches with high estimates of generalization error, based on the training data.


\subsection{Split search} \label{split-search}

In top-down \ac{DT} learning, suppose that a new node~$\mathcal{n}$ sees the training subset $\mathcal{D}^{(\mathcal{n})}\subset\mathcal{D}$. In a binary \ac{DT}, we define the branch function at $\mathcal{n}$ as $B^{(\mathcal{n})}(\mathbf{x})$ that indicates if the instance $\mathbf{x}$ goes to the left or right child of~$\mathcal{n}$, $\mathcal{n}_L$ or $\mathcal{n}_R$. $B^{(\mathcal{n})}(\mathbf{x})$ is parametrized by the attribute index $j^{(\mathcal{n})}$ and the threshold~$\tau^{(\mathcal{n})}$:
\begin{equation}
B^{(\mathcal{n})}(\mathbf{x}) = 
\left\{
\begin{array}{ll}
\mathcal{n}_L & \mbox{if  } x_{j^{(\mathcal{n})}} < \tau^{(\mathcal{n})} \\
\mathcal{n}_R & \mbox{otherwise},
\end{array}
\right.
\label{eqn:branch}
\end{equation}

\noindent
where $x_{j^{(\mathcal{n})}}$ is the observation of variable $X_{j^{(\mathcal{n})}}$ for~$\mathbf{x}$. The split search consists in finding the attribute $X_j$ and threshold $\tau$ that split $\mathcal{D}^{(\mathcal{n})}$ with maximum class separation.

The entropy of variable~$Y$ is defined as:
\[
H_\alpha(Y) = - \sum_{y} p(y) {log}_\alpha p(y),
\]
\noindent
We take $\alpha=2$ such that the entropy is measured in \textit{bit}, and omit~$\alpha$. In \ac{ID3}, increasing class purity corresponds to reducing $H(Y)$ in $\mathcal{n}_L$ and $\mathcal{n}_R$, compared to~$\mathcal{n}$. In other words, $j^{(\mathcal{n})}$ and $\tau^{(\mathcal{n})}$ are chosen to maximize the \textit{Mutual Information} between $B^{(\mathcal{n})}$ and~$Y$,~$I(B^{(\mathcal{n})};Y)$:
\begin{equation}
j^{(\mathcal{n})}, \tau^{(\mathcal{n})} = \operatorname*{arg\,max}_{j, \tau} I(B^{(\mathcal{n})};Y).
\label{eqn:infoGain}
\end{equation}
\noindent 
In the \ac{DT} literature, $I(B^{(\mathcal{n})};Y)$ is known as \textit{information gain}, and is equal to the difference between the entropy of~$Y$ in~$\mathcal{D}^{(\mathcal{n})}$ and the entropy of~$Y$ in the resulting nodes:
\[
I(B^{(\mathcal{n})};Y) = H(Y) - H(Y|B^{(\mathcal{n})}).
\]
\noindent
The term $H(Y|B^{(\mathcal{n})})$ is the conditional entropy of $Y$ given the split variable~$B^{(\mathcal{n})}$. Equation~\ref{eqn:infoGain} is equivalent to:
\begin{align}
j^{(\mathcal{n})},\tau^{(\mathcal{n})}		& = \operatorname*{arg\,min}_{j, \tau} H(Y|B^{(\mathcal{n})}) \nonumber \\
& = \operatorname*{arg\,min}_{j, \tau} \sum_{b \in \{\mathcal{n}_L ,\mathcal{n}_R\}}{p(b)H(Y|B^{(\mathcal{n})}=b)},
\label{eqn:condEnt}
\end{align}
\noindent where
\begin{equation}
H(Y|B^{(\mathcal{n})}=b)	 = - \sum_{y=1}^{K}{ p(y|b)log_2p(y|b) }.
\label{eqn:condEnt2}
\end{equation}
\noindent
The maximum-likelihood estimates of the probabilities in Equations~\ref{eqn:condEnt} and~\ref{eqn:condEnt2} from the training data are:
\begin{gather}
\hat{p}(b) = \tfrac{N_\mathcal{n}(b)}{N_\mathcal{n}} \text{ and } \hat{p}(y|b) = \tfrac{N_\mathcal{n}(y,b)}{N_\mathcal{n}(b)},
\label{eqn:probs}
\end{gather}

\noindent where $N_\mathcal{n}$ is the number of training instances in~$\mathcal{n}$, $N_\mathcal{n} = \sum_{\mathbf{x} \in \mathcal{D}^{(\mathcal{n})}} 1$, and $N_\mathcal{n}(e)$ is the number of instances in $\mathcal{n}$ for which event $e$ was observed, $N_\mathcal{n}(e) = \sum_{\mathbf{x} \in \mathcal{D}^{(\mathcal{n})}} \mathbf{1}(e)$.

In this manuscript, we consider numerical variables, and employ binary splits and use the information gain as a split criterion. In C4.5 release 8, the best split for each numerical attribute is first selected using information gain. Subsequently, the \textit{gain ratio} criterion is used to compare the best splits found for all variables, categorical and numerical~\cite{Quinlanpersonal}.

\footnote{C4.5 performs $n$-ary splits of categorical variables. Since the information gain is biased towards splits with many outcomes, C4.5 normalizes it by the entropy of the $B^{(\mathcal{n})}$ variable, defining the \textit{gain ratio}. In C4.5 release 8, the best split for each numerical attribute is first selected using information gain. Subsequently, the gain ratio is used to compare the best splits found for all variables, categorical and numerical\cite{Quinlanpersonal}.}

\begin{equation}
\tfrac {H(Y|B_j)} {H(B_j)}.
\end{equation}


\subsection{Probabilistic interpretation of the splits} \label{probabilistic}

Consider $p(y|\mathbf{x})$ the probability of $y$ given instance $\mathbf{x}$, based on which we can make a prediction about $y$. We can estimate $p(y|\mathbf{x})$ from the \ac{DT} rooted in node $\mathcal{n}$, with child nodes $\mathcal{n}_L$ and~$\mathcal{n}_R$ as:
\begin{align}
\begin{split}	
\hat{p}(y|\mathbf{x}) & = \textstyle \sum_{b \in \{ \mathcal{n}_L, \mathcal{n}_R \}}{ p(y|b, \mathbf{x}) p(b|\mathbf{x}) } \\
& = \textstyle \sum_{b \in \{ \mathcal{n}_L, \mathcal{n}_R \}}{ p(y|b) p(b|\mathbf{x}) } \\
& = p(y|\mathcal{n}_L) p(\mathcal{n}_L|\mathbf{x}) + p(y|\mathcal{n}_R) p(\mathcal{n}_R|\mathbf{x}),
\label{eqn:splits}
\end{split}
\end{align}

\noindent
where $y$ and $\mathbf{x}$ are conditionally independent given~$B^{(n)}=b$. Equation~\ref{eqn:splits} is based on Bayesian model averaging~\cite{hoeting1999bayesian}, which translates the uncertainty of the distinct models, in this case the two subtrees $\mathcal{n}_L$ and $\mathcal{n}_R$, into uncertainty in the class prediction.

Probabilistic \acp{DT} use this idea to instead express the uncertainty about the observed instance~$\mathbf{x}$. This uncertainty is modeled by the posterior $p(\mathcal{n}_L|\mathbf{x})$, known as the \textit{gating function}, $g_\mathcal{n}(\mathbf{x})$~\cite{irsoy2012soft}. The estimate of $p(y|\mathbf{x})$ becomes:
\begin{equation}
\hat{p}(y|\mathbf{x}) = p(y|\mathcal{n}_L) g_\mathcal{n}(\mathbf{x}) + p(y|\mathcal{n}_R) \big(1 - g_n(\mathbf{x})\big).
\label{eqn:prediction}
\end{equation}

If we assume that the observed value is accurate, node $\mathcal{n}$ performs a \textit{hard split} as in Equation~\ref{eqn:branch}, and so:
\[
g_\mathcal{n}(\mathbf{x}) = \mathbf{1}(x_{j^{(\mathcal{n})}} < \tau^{(\mathcal{n})}),
\] 
\noindent In this case, if $x_{j^{(\mathcal{n})}}$ is close to $\tau^{(\mathcal{n})}$, small variations in its value can drastically change the estimate $p(y|\mathbf{x})$, and produce incorrect predictions~\cite{Quinlan1987}. Probabilistic \acp{DT} instead model the uncertainty each variable as a distribution of noise, and $g_\mathcal{n}(\mathbf{x})$ becomes the \ac{CDF} of the chosen distribution. A small variation of $x_{j^{(\mathcal{n})}}$ around the threshold value will then smoothly alter~$p(y|\mathbf{x})$.

When searching for the best split for a node $\mathcal{n}$, we use the training subset~$\mathcal{D}^{(\mathcal{n})}$ to obtain the probability estimates of Equation \ref{eqn:probs}. E.g. we can estimate $p(\mathcal{n}_L)$, or equivalently $p(\mathcal{n}_R)$, in terms of~$g_\mathcal{n}(\mathbf{x})$:
\begin{align}
\begin{split}
\hat{p}(\mathcal{n}_L) & = \tfrac{ \sum_{\mathbf{x} \in \mathcal{D}^{(\mathcal{n})}} p(\mathcal{n}_L|\mathbf{x})}{\sum_{\mathbf{x} \in \mathcal{D}^{(\mathcal{n})}}1}
= \tfrac{ \sum_{\mathbf{x} \in \mathcal{D}^{(\mathcal{n})}} g_\mathcal{n}(\mathbf{x})}{\sum_{\mathbf{x} \in \mathcal{D}^{(\mathcal{n})}}1}
\end{split}
\label{eqn:splits-train}
\end{align}

\section{Proposed approach} \label{proposed-approach}

We propose a probabilistic \ac{DT} approach to handle uncertainty by modeling it as a distribution of noise around the observed value. The noise model should expresses existing knowledge about the uncertainty. This model is independently considered:
\begin{enumerate}
	\item When searching for the split thresholds, during training, or  (Section ~\ref{soft-search}, \acf{SS}),
	\item When propagating the training instances, during training (Section ~\ref{soft-propagation-training}, \acf{STP}),
	\item When propagating test instances through the constructed~\ac{DT}, during evaluation \ac{DT} (Section ~\ref{soft-evaluation}, \acf{SE}).
\end{enumerate}
As a proof-of-concept, we consider that the noise of variable~$X$ is additive and captured by the variable $\mathcal{E}_{X} \sim \mathcal{N}(0, \sigma^2)$. We consider the normal distribution to be a good assumption to study our approach. It can be useful for various types of data, owing to the Central Limit Theorem, and is fully described by its mean and~variance.

\subsection{Soft search} \label{soft-search}

Let us focus on the computation of the information gain for variable~$X$. Suppose that $X$ takes $D$ \textit{distinct} values sorted as $\{ x^{(1)}, ..., x^{(D)} \}$ with $x^{(i)} < x^{(i+1)}$, $i=1, ..., D-1$ in the training data $\mathcal{D}$. Note that $|\mathcal{D}| \ge D$. We focus on finding a split for node~$\mathcal{n}$, and consider that all the concepts in this Section refer to~$\mathcal{n}$, and omit the subscript. $N(x^{(i)})$~is the number of instances with value $x^{(i)}$, and $N(y, x^{(i)})$ the number of such instances that belong to class~$y$. Let $\tau$ denote the candidate threshold for a split based on $X$ at node~$\mathcal{n}$. 



In C4.5, the split search is done by computing the information gain in Equation~\ref{eqn:infoGain} for each candidate $\tau = x^{(i)}$, $i=2, ..., D$. Since the method assumes certain measurements, we have $g(\mathbf{x}) = \mathbf{1}(x<\tau)$. The probabilities $p(b)$ and $p(y|b)$ are estimated as in Equation~\ref{eqn:probs}. We now describe how to estimate these probabilities efficiently, considering uncertain measurements.  





If we consider the noise model $\mathcal{E}_{X} \sim \mathcal{N}(0, \sigma^2)$, the gating function will be the normal~\ac{CDF}. Let us denote the numerator of Equation~\ref{eqn:splits-train} as $\rho(\tau)$, representing the density of training examples falling on the left child node:
\begin{align}
\begin{split}
\rho(\tau) = \sum_{\mathbf{x} \in \mathcal{D}} p(\mathcal{n}_L|\mathbf{x}) =	
\sum_{\mathbf{x} \in \mathcal{D}} g_\mathcal{n}(\mathbf{x}) 
=  \sum_{\mathbf{x} \in \mathcal{D}} \int_{-\infty}^{\tau} K(\tau - x, \sigma) d\tau,
\end{split}
\end{align}
\noindent where $K(\tau - x, \sigma)$ is the Gaussian kernel function centered on the measurement $x$ with variance~$\sigma^2$:
\begin{equation}
K(\tau - x, \sigma) \propto exp(-\tfrac{(\tau - x)^2}{2\sigma^2}).
\end{equation}
\noindent This corresponds to using Gaussian kernel density estimation for the probabilities used to compute the information gain. Since $\rho(\tau)$ is no longer constant between each $x^{(i)}$ and $x^{(i+1)}$, the candidate thresholds need not be the dataset values. We take:
\begin{equation}
\tau = \tau_{min}, ~ \tau_{min} + \delta \sigma, ~ \tau_{min} + 2 \delta \sigma, ~ ..., ~ \tau_{max}
\label{eqn:discretization}
\end{equation}
\noindent
with $\tau_{min} =  x^{(1)}-\tfrac{\omega}{2}\sigma$ and $\tau_{max} \le x^{(D)}+\tfrac{\omega}{2}\sigma$. The parameter $\delta$ controls the resolution of the search, and $\omega$ is the \textit{window factor}, $\omega>\delta$. For efficiency, we consider the kernel adjusted to be zero for~$|\tau - x| > \omega$. Using Equation~\ref{eqn:discretization}, the number of information gain computations is limited to $\tfrac{1}{\delta} ( \tau_{max} - \tau_{min} )$.

To compute the information gain efficiently for the candidate splits, we keep two running sums of $\rho(\tau)$, on the left and on the right of~$\tau$,  $\rho_L(\tau)$ and  $\rho_R(\tau)$. We initialize the search with  $\rho_L(\tau) = 0$ and  $\rho_R(\tau)=N$. Each time $\tau$ is incremented $\delta\sigma$, the left and right sums are respectively incremented and decremented the quantity $\Delta\rho(\tau)$:
\begin{equation} 
\Delta\rho(\tau) = 
\begin{cases} 
\mathlarger{\sum}_{i=1}^{D}{N(x^{(i)}) \Phi \big(\tfrac{\tau - x^{(i)}}{\sigma} \big) },
\hfill \mbox{ ~~if } \tau=\tau_{min} \\
\mathlarger{\sum}_{i=1}^{D}{N(x^{(i)}) \Big[ \Phi \big(\tfrac{\tau - x^{(i)}}{\sigma} \big) - \Phi \big(\tfrac{\tau - \delta \sigma - x^{(i)} }{\sigma} \big)\Big] }, 	\\
\hfill  \mbox{ ~~if } \tau_{min} < \tau < \tau_{max} \\
\mathlarger{\sum}_{i=1}^{D}{N(x^{(i)}) \Big[ 1 - \Phi \big(\tfrac{\tau - \delta \sigma - x^{(i)} }{\sigma} \big)\Big] }, 	
\hfill  \mbox{ ~~if } \tau=\tau_{max} 
\end{cases}
\label{eqn:filter}
\end{equation}
\noindent with $\Phi(\cdot)$ the \ac{CDF} of the standard normal distribution. The contribution of measurement $x^{(i)}$ to $\Delta\rho(\tau)$ is proportional to $\Phi \big(\tfrac{\tau - x^{(i)}}{\sigma} \big) - \Phi \big(\tfrac{\tau - x^{(i)} - \delta \sigma}{\sigma} \big)$. The last point $\tau_{max}$ ensures that the density contributions of $x^{(i)}$ sum to~$N(x^{(i)})$.


Similarly, to estimate $\hat{p}(y|{\mathcal{n}}_L)$, we consider the sum of the densities per class, $\rho(y,\tau)$. The density increments per class $\Delta\rho(y,\tau)$ are computed by replacing the number of instances $N(x^{(i)})$ by $N(y, x^{(i)})$ in Equation~\ref{eqn:filter}. Finally, the estimated probabilities $\hat{p}({\mathcal{n}}_L)$ and $\hat{p}(y|{\mathcal{n}}_L)$ are used to minimize the conditional entropy $H(Y|B)$ in Equation~\ref{eqn:condEnt}. Searching for the threshold using the set of values in Equation~\ref{eqn:discretization} and the density increments $\Delta\rho(\tau)$ and $\Delta\rho(y,\tau)$ acts as a Gaussian filter to the information gain. We refer to this approach as \acf{SS}.



The standard deviation of $\mathcal{E}_{X}$ is set to $\sigma=u_{s} \bar x$, where $u_{s}$ is the \ac{SS} \textit{uncertainty factor} and $\bar x$ is the sample mean of~$X$. This choice of $\sigma$ was made, as clinicians often report the uncertainty in terms of the absolute values of the attributes~\cite{d2016two}. In the following experiments, $u_s$ is assumed to be the same for all variables, although it could be specified independently.





\subsection{Soft training propagation} \label{soft-propagation-training}

We can also account for the uncertainty when splitting the training data, which we call \acf{STP}. The two soft training approaches, \ac{SS} and \ac{STP} can be used in combination or independently.

Consider the split variable $X$ and value~$\tau^{(\mathcal{n})}$ at node~$\mathcal{n}$. As before, a soft split is achieved by setting the gating function to the \ac{CDF} of~$\mathcal{E}_{X}$ centered in $x$:
\begin{equation}
g_{\cal{n}}(\mathbf{x}) = F_{\mathcal{E}_{X}}(\tau - x) = \Phi \big(\tfrac{\tau - x}{\sigma} \big).
\label{eqn:soft-splits}
\end{equation}
\noindent with $F_{\mathcal{E}_{X}}(.)$ the \ac{CDF} of~$\mathcal{E}_{X}$. Each training instance is fractionally divided between the child nodes, according to the probability~$g_{\cal{n}}(\mathbf{x})$. The standard deviation is set to $\sigma=u_{p} \bar x$, with $u_{p}$ the \ac{STP} uncertainty factor, and $\bar x$ the variable mean.

\Ac{STP} leads to increased learning times. Using hard splits, each instance is sent down one branch, and the total number of instances remains constant at each level of the tree. The number of information gain computations at any \ac{DT} depth is bounded by the dataset size and the number of attributes, $|\mathcal{D}| M$. With \ac{STP}, each instance is sent down all the branches, with weights determined by the gating function. Therefore, a node sees a \textit{greater} number of instances compared to the hard approach. The number of information gain computations at a given \ac{DT} depth $d$ is raised to~$2^{d} |\mathcal{D}| M$.


\subsection{Soft evaluation} \label{soft-evaluation}

The uncertainty may also be accounted for when classifying test cases with a finalized~\ac{DT}. \Acf{SE} is achieved by setting the gating function as in Equation~\ref{eqn:soft-splits}, but with $\sigma=u_{e} \bar x$. The evaluation uncertainty factor is denoted as~$u_e$ and $\bar x$ is obtained from the \textit{training} data. From Equation~\ref{eqn:prediction}, the class probability estimates for instance~$\mathbf{x}$ become:
\begin{equation}
\hat{p}(y|\mathbf{x}) = p(y| \mathcal{n}_L)\Phi(\tfrac{\tau-x}{\sigma}) + p(y|\mathcal{n}_R)\big(1 - \Phi(\tfrac{\tau-x}{\sigma}) \big).
\label{eqn:comp2}
\end{equation}

\noindent As described in Section~\ref{probabilistic}, this approach adjusts $\hat{p}(y|\mathbf{x})$ to reflect the choice of noise distribution, in this case, Gaussian.

\subsection{Motivation on a toy example} \label{motivation}

\noindent To better motivate the use of the soft algorithm components, we consider a toy example with a input variable $X$ and class~$Y \in \{0, 1\}$.

\subsubsection*{\Acl{SS}}

\begin{figure*}	\centering
	\begin{subfigure}[b]{\textwidth} \centering
		\includegraphics[width=\textwidth]{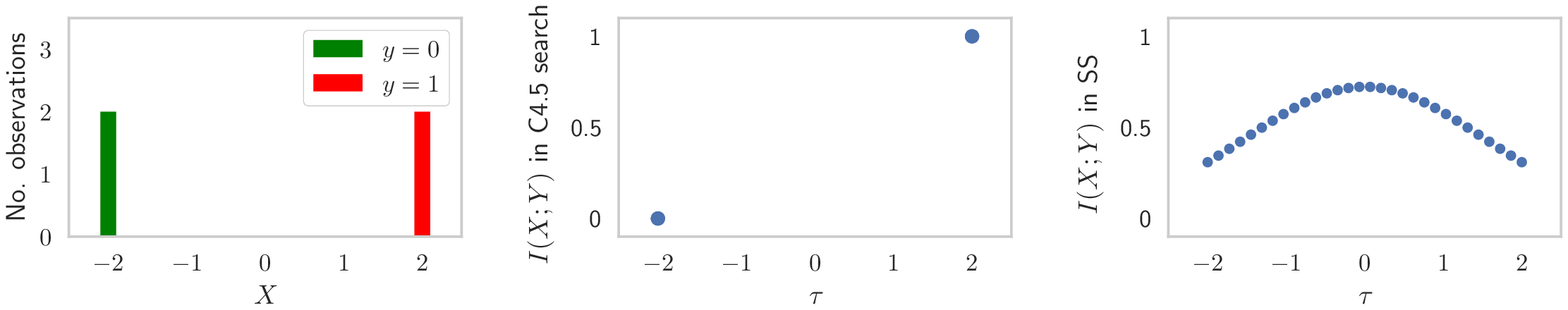} \caption{Example with $(x_{(1)}, y_{(1)}) = (x_{(2)}, y_{(2)})=(-2,0)$ and $(x_{(3)}, y_{(3)})=(x_{(4)}, y_{(4)})=(2, 1)$.\\} \label{fig:4point-a}
	\end{subfigure}
	\begin{subfigure}[b]{\textwidth} \centering
		\includegraphics[width=\textwidth]{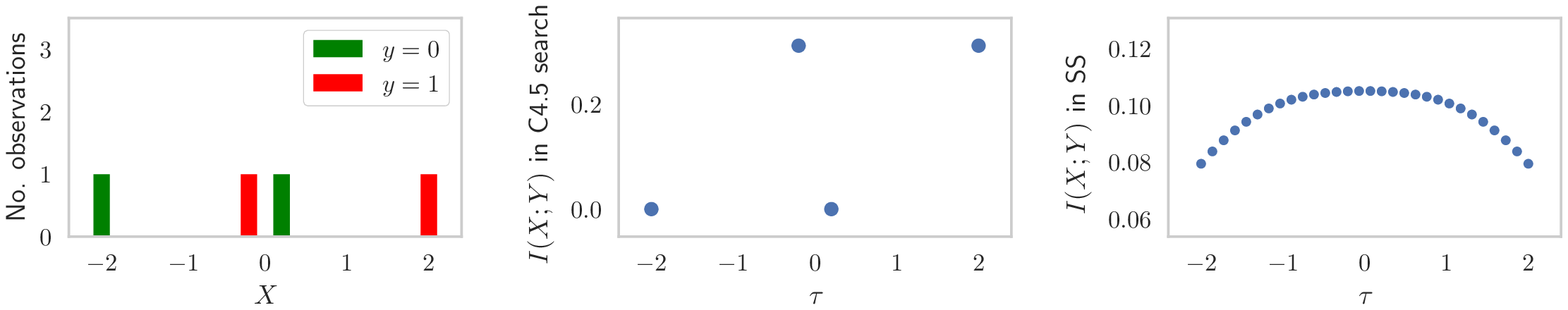} \caption{Example with $(x_{(1)}, y_{(1)}) = (-2,0)$ and $(x_{(4)}, y_{(4)})=(2, 1)$, and noisy measurements $(x_{(2)}, y_{(2)})=(0.2,0)$ and $(x_{(3)}, y_{(3)})= (-0.2, 1)$.} \label{fig:4point-b}
	\end{subfigure}
	\caption{Motivating example to show the effect of employing the \acf{SS} on the information gain $I(X;Y)$, with $\sigma_s=0.3$.}
	\label{fig:4point}
\end{figure*}

Suppose we have a training dataset of four examples $\{ (x_{(1)} ,0),  (x_{(2)} ,0), (x_{(3)} ,1), (x_{(4)} ,1)\}$, with $x_{(1)} = x_{(2)} = -2$ and $x_{(3)} = x_{(4)} = 2$. Figure~\ref{fig:4point-a} displays the information gain $I(X;Y)$ computed by C4.5 and the \ac{SS} for the corresponding candidate splits, denoted by~$\tau$. C4.5 would choose $\tau = 2$ as a split, while \ac{SS} would choose a split close to $\tau = 0$, potentially avoiding the misclassification of test examples in the interval $(0,2)$, for which there is no training data. If the measurements are noisy such that e.g. $x_{(2)}=0.2$ and $x_{(3)}=-0.2$, C4.5 would select $\tau=0.2$ or $\tau=2$, as seen in Figure~\ref{fig:4point-b}. The \ac{SS} would select a value close to zero.

\subsubsection*{\Acl{STP}}

Let us now discard two training examples, and keep $\mathbf{x}_{(1)} = (-2,0)$ and $\mathbf{x}_{(4)} = (2,1)$. We model the uncertainty of $X$, such that $x_{(1)} \sim \mathcal{N}(-2, \sigma^2)$ and $x_{(4)} \sim \mathcal{N}(2, \sigma^2)$ are drawn from normal distributions with mean $\mu_{(1)} = -2$ and $\mu_{(2)} = 2$, and common variance $\sigma^2$. Let us also consider the test point $\mathbf{x}_{(t)}=(x_{(t)}, 1)$. We analyze the probability of misclassifying $\mathbf{x}_{(t)}$. We note that the sum and difference of normal distributions are also normally-distributed:
\begin{gather*}
d = x_{(1)} - x_{(4)} \sim \mathcal{N}(-2 -2, \sigma^2 + \sigma^2 ) = \mathcal{N}(-4, 2\sigma^2 ),\\
s = x_{(1)} + x_{(4)} \sim \mathcal{N}(-2 + 2, \sigma^2 + \sigma^2 ) = \mathcal{N}(0, 2\sigma^2 ), \\
m = \tfrac{s}{2} =\tfrac{x_{(1)} + x_{(4)}}{2} \sim \mathcal{N}(\tfrac{1}{2} \cdot 0, \tfrac{1}{2^2}) = \mathcal{N}(0, \tfrac{\sigma^2}{2}) 
\end{gather*}
\noindent Let us assume that the search selects the midpoint $ m$ as a split value. The prediction $\hat{y}(x_{(t)})$ depends on whether the difference $c=x_{(t)}-m$ is greater than or less than~$0$. Specifically, the probability of misclassifying $\mathbf{x}_{(t)}$ is composed by two terms:

\begin{figure}[h]
	\captionsetup[subfigure]{justification=centering} \centering
	\begin{subfigure}{.49\textwidth} \centering 
		\includegraphics[width=.95\textwidth]{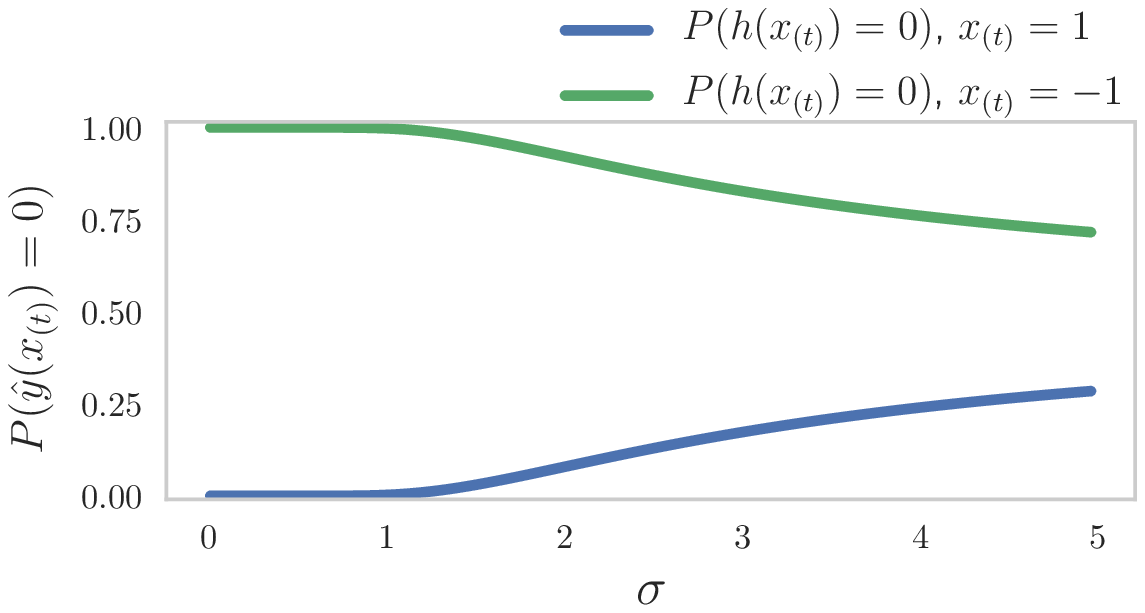} \vspace{-4mm}
		\caption{ $P\big(\hat{y}(x_{(t)})$ when we assume $x_{(t)}$ to be certain.} \label{fig:perror-ex1a}
	\end{subfigure}
	\begin{subfigure}{.49\textwidth} \centering 
		\includegraphics[width=.95\textwidth]{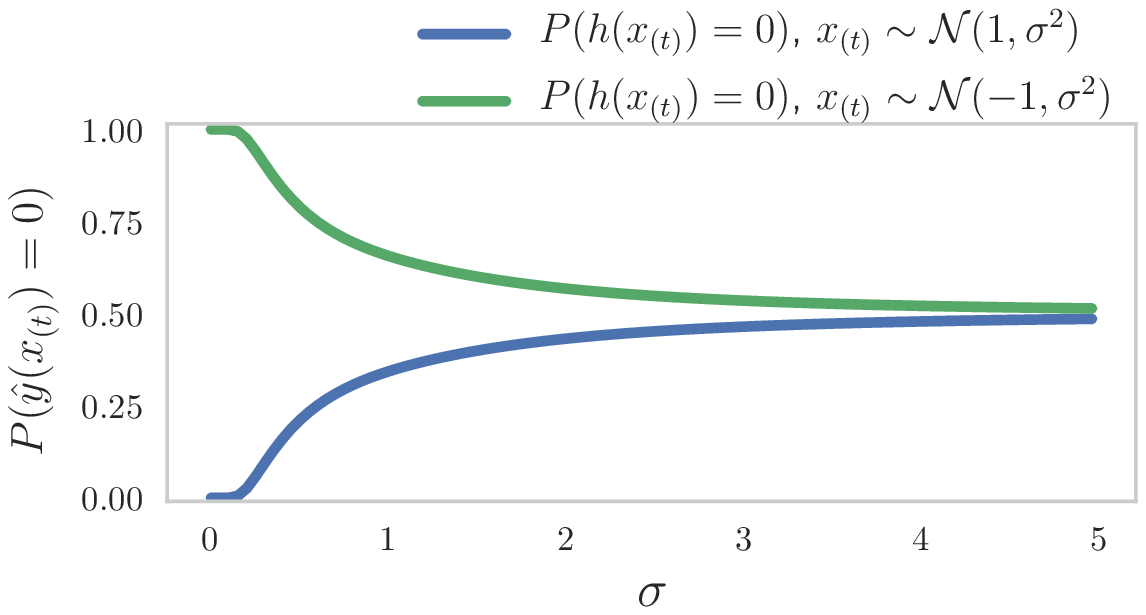} \vspace{-4mm}
		\caption{ $P\big(\hat{y}(x_{(t)})$ when $x_{(t)}$ is assumed to have noise.} \label{fig:perror-ex1b}
	\end{subfigure}
	\caption{Probability of misclassifying $(x_{(t)}, 1)$ as function of the standard deviation $\sigma$ of the normal uncertainty model. In \ref{fig:perror-ex1a}, the uncertainty model is considered only for the training instances $x_{(1)}$ and $x_{(4)}$, simulating \acf{STP}, while $x_{(t)}$ is certain. In \ref{fig:perror-ex1b} $x_{(t)}$ has normally-distributed noise, as in \acf{SE}.}
	\label{fig:perror-ex1}
\end{figure}

\begin{gather*}
P\big(\hat{y}(x_{(t)}) = 0\big) = P (d < 0)P (c < 0) + P (d > 0)P (c > 0).
\end{gather*}
\noindent In case $x_{(1)}$ is smaller than~$x_{(4)}$, C4.5 assigns class $0$ to $x_{(t)}$ iff~$x_{(t)} < m$. In case $x_{(1)}$ is greater than~$x_{(4)}$, C4.5 assigns class $0$ to $x_{(t)}$ iff~$m < x_{(t)}$. Each of the above probabilities is given by the \ac{CDF} associated with the corresponding distribution. If we shift each distribution to have $0$ mean, we can express the probabilities as:
\begin{gather*}
P (d < 0) = P (d + 4 < 4) = F_{d+4}(4) = \Phi\big(\tfrac{4}{2 \sigma}\big), \\
P (d > 0) = 1 - P (d < 0) = 1 - \Phi\big(\tfrac{4}{2 \sigma}\big).
\end{gather*}
$F_{d+4}$ is the \ac{CDF} of $d+4$. Figure~\ref{fig:perror-ex1a} displays the probability of misclassifying $x_{(t)}$ as a function of $\sigma$ when $x_{(t)}$ is certain. When $x_{(t)} = 1$, and so $c > 0$, $P\big(\hat{y}(x_{(t)}) = 0\big)$ is nearly zero until the distributions of the training instances start to overlap with increasing~$\sigma$. The inverse occurs for $c < 0$, where the error probability starts to decrease. Figure~\ref{fig:perror-ex1a} shows how the prediction probabilities change as the model expresses less confidence on the training data, when using~\ac{STP}.

\subsubsection*{\Acl{SE}}

Let us now express the uncertainty about the test example, such that $x_{(t)}$ is also normally distributed with variance~$\sigma^2$. We can obtain $P (c < 0)$ and $P (c > 0)$ as:
\begin{align*}
\begin{split}
P (c < 0) = P ( c - 1 < -1 ) = F_{c-1}(-1) & = 1 - F_{c-1}(1) = \\
& = 1 - \Phi\big(\tfrac{\sqrt{2}}{\sqrt{3} \sigma}\big), \\
P (c > 0) = 1 - P ( c < 0 ) = \Phi\big(\tfrac{\sqrt{2}}{\sqrt{3} \sigma}\big).
\end{split}	
\end{align*}

\noindent Figure~\ref{fig:perror-ex1b} displays $P\big(\hat{y}(x_{(t)}=0)$ for $x_{(t)} \sim \mathcal{N}(1, \sigma^2)$ and $x_{(t)} \sim \mathcal{N}(-1, \sigma^22)$, which now converges faster to~$0.5$.

\section{Experiment design}

The experiments assess effect of the proposed uncertainty model in the distinct \ac{DT} learning phases, and the impact of corrupting the training versus the test data with noise.

Improving prediction performance consists in maximizing the generalization accuracy, i.e. the fraction of correctly classified test examples, while minimizing model complexity~\cite{Kearns1997}. \Ac{DT} complexity is assessed by measuring its number of leaves.

Each proposed soft component, \ac{SS}, \ac{STP} and \ac{SE} is independently compared to C4.5. The C4.5 pruning and missing-value strategies are equally employed in all experiments. Pruning is extended with the Laplace correction, as recommended to favor the exploration of smaller \acp{DT} for the same accuracy~\cite{Niblett1986}. We also extend the evaluation of the \ac{PLT}~\cite{quinlan1990probabilistic} and \ac{UDT}~\cite{Tsang2011} approaches to more data and noise scenarios.


\subsection{Data description and pruning confidence factor} \label{data}

Table~\ref{tab:datasets} displays the employed datasets. We sought clinical data from open resources, including UCI \ac{ML}~\cite{Lichman2013} and KEEL~\cite{Alcala-Fdez2011} repositories. Non-ordinal features were excluded. We synthesized $5$ additional datasets using an adaptation of the method by Guyon \cite{Guyon2003}, available in Scikit-learn's implementation \textit{make\_classification}~\cite{scikit-learn}. The datasets were generated with varying properties.

Since optimal \ac{DT} size is problem- and dataset-dependent~\cite{jensen1999effects}, the pruning confidence factor is not optimized. Instead, for each dataset the confidence factor is fixed such that the average \ac{DT} size achieved by C4.5 through \ac{CV} is 15 leaves. This corresponds to a binary tree wit nearly $4$ levels, considered a manageable/interpretable number of decisions in a visual clinical guideline. Some of the datasets were too small to reach 15 leaves, so their confidence factor is set to either 10 or 5 leaves. The confidence factor is fixed across all experiments.

\subsection{Experiments} \label{model-evaluation}

For each real dataset, $30$ random train-test permutations were created, containing respectively 70\% and 30\% of the data. Stratified sampling was used, so that class proportions are equal in all samples. For each synthetic dataset, distinct instances were generated and divided into 30 different sets, which were then split into 70\%-30\% train-test samples.

The experiments are performed with varying degrees of noise to the data. The noise added to a variable $X$ in a data subset is sampled according to $\mathcal{N}(0,n \bar x)$, with $n$ the \textit{noise factor} and $\bar x$ the training subset mean of $X$. The same $n$ is used for all its variables. All randomness was generated with fixed seeds for reproducibility.

\begin{table}
	\renewcommand{\arraystretch}{0.97} \centering
	\caption{ Classification datasets used in the experiments. }
	\label{tab:datasets}
	\setlength{\tabcolsep}{3.2pt}
	\begin{tabular}{llccc}
		\toprule
		\multirow{2}{*}{\specialcell{Dataset}} & \multirow{2}{*}{\specialcell{Source}}								& \multirow{2}{*}{\specialcell{No.\\variables}}	& \multirow{2}{*}{\specialcell{No.\\instances}}			& \multirow{2}{*}{\specialcell{No.\\classes}} \\ \\
		\midrule \midrule
		Hepatitis							& KEEL				& 6			& 155				& 2		\\
		Heart disease						& UCI \ac{ML}		& 8			& 303				& 2		\\
		Pima Indians diabetes				& UCI \ac{ML}		& 8			& 768				& 2		\\
		South African heart					& UCI \ac{ML}		& 8			& 462				& 2		\\		
		Breast cancer Wisconsin				& UCI \ac{ML}		& 39		& 569				& 2		\\		
		Dermatology							& UCI \ac{ML}		& 34		& 366				& 6		\\
		Haberman's breast cancer			& UCI \ac{ML}		& 3			& 306				& 2		\\
		Indian liver patient 				& UCI \ac{ML}		& 9			& 582				& 2		\\
		BUPA liver disorders				& UCI \ac{ML}		& 6			& 345				& 2		\\
		Vertebral column (2 classes)		& UCI \ac{ML}		& 12		& 310				& 2		\\
		Vertebral column (3 classes)		& UCI \ac{ML}		& 12		& 310				& 3		\\
		Thyroid gland						& UCI \ac{ML}		& 5			& 215				& 3		\\
		Oxford Parkinson's disease 			& UCI \ac{ML}		& 22		& 194				& 2		\\
		SPECTF								& UCI \ac{ML}		& 44		& 266				& 2		\\
		Thoracic surgery					& UCI \ac{ML}		& 3			& 470				& 2		\\
		Synthetic 1							& Guyon							& 15			& 30$\times$500		& 2		\\
		Synthetic 2							& Guyon							& 15			& 30$\times$400		& 2		\\
		Synthetic 3							& Guyon							& 20			& 30$\times$300		& 2		\\
		Synthetic 4							& Guyon							& 25			& 30$\times$200		& 3		\\
		Synthetic 5							& Guyon							& 20			& 30$\times$250		& 3		\\ 
		\bottomrule
	\end{tabular}
\end{table}

\subsubsection{Experiment 1: noise added to the training data}\label{experiment-1}

Experiment~$1$ studies the approaches with noise added to the training data, so we denote the training noise factor as~$n_{train}$. The uncertainty factors $u_s$, $u_t$, $u_e$ of the \ac{SS}, \ac{STP} and \ac{SE} approaches, as well as the \ac{UDT} parameter $w$, control the standard deviation of the uncertainty model. As such, we tune them by \acf{CV} for each dataset and~$n_{train}$. \noindent The \ac{SS} parameters $\omega$ and $\delta$ are respectively set to $6.0$ and $0.1$. Initial experiments showed that they do not significantly impact the results, provided that $\omega$ is large enough to contain most of the density of the uncertainty distribution, and $\delta$ is small enough to ensure a large set of candidate splits. The \ac{PLT} parameters $\tau^-$ and $\tau^+$ are derived using the method proposed by Quinlan~\cite{quinlan1990probabilistic}. We summarize the steps for model tuning and evaluation:
\begin{quote}
	\begin{small}
		For each $70\%-30\%$ train-test permutation, $n_{train}$ and model: \\
		1. Hold out the 30\% test set \\
		2. If model has parameter to set ($u_s$, $u_t$, $u_e$, or $w$), use the 70\% training set to tune it by \ac{CV}: \\
		\text{~~~~~}(a) Compute 10 stratified \ac{CV} folds. \\
		\text{~~~~~}(b) Add noise to the \ac{CV} training folds, distributed \\ \text{~~~~~~~~~} as $\mathcal{N}(0,n_{train} \bar x)$. Do not add noise to the \ac{CV} \\ \text{~~~~~~~~~} validation folds. \\
		\text{~~~~~}(c) Tune the parameter to maximize \ac{CV} accuracy.\\
		3. Add noise to the initial 70\% training set, distributed as $\mathcal{N}(0,n_{train} \bar x)$. \\
		4. Learn a tree using the noisy training set, and the selected parameter value, if applicable. \\
		5. Evaluate the tree on the 30\% test set, to which no noise was added.
	\end{small}
\end{quote}

\noindent C4.5 and \ac{PLT} do not undergo step $2$.

\subsubsection{Experiment 2: noise added to the test data}\label{experiment-2}

Experiment~$2$ evaluates the merit of the models built on data without added noise in predicting the labels of noisy test cases. Noise is added to the data used for evaluation, and we denote $n$ as~$n_{test}$. Model evaluation is done as:
\begin{quote}\begin{small}
		For each $70\%-30\%$ train-test permutation, $n_{test}$ and model: \\
		1. Hold out the 30\% test set \\
		2. If model has parameter to set ($u_s$, $u_t$, $u_e$, or $w$), use the 70\% training set to tune it by \ac{CV}: \\
		\text{~~~~~}(a) Compute 10 stratified \ac{CV} folds. \\
		\text{~~~~~}(b) Do not add noise to the \ac{CV} training folds. \\ \text{~~~~~~~~~} Add noise to the \ac{CV} validation folds, distri-\\ \text{~~~~~~~~~~}buted as $\mathcal{N}(0,n_{test} \bar x)$. \\
		\text{~~~~~}(c) Tune the parameter to maximize \ac{CV} accuracy.\\
		3. Learn a tree using the initial 70\% training set, and the selected parameter value, if applicable \\
		4. Add noise to the 30\% test set, distributed as $\mathcal{N}(0,n_{test} \bar x)$. \\
		5. Evaluate the tree on noisy the 30\% test set.
\end{small}\end{quote}

\noindent As before, C4.5 and \ac{PLT} do not undergo step 2. Note that noise is added to the \ac{CV} validation folds.

\section{Results} \label{results}

Section~\ref{ef-example} illustrates \ac{SS}, \ac{STP} and \ac{SE} on a single variable. In Section~\ref{exp-res}, we show the results of Experiments~$1$ and $2$.

\subsection{Illustration on a single variable} \label{ef-example}

\begin{figure}[t]
	\captionsetup[subfigure]{justification=centering}
	\centering
	\begin{subfigure}{.48\textwidth} \centering
		\includegraphics[height=4.3cm]{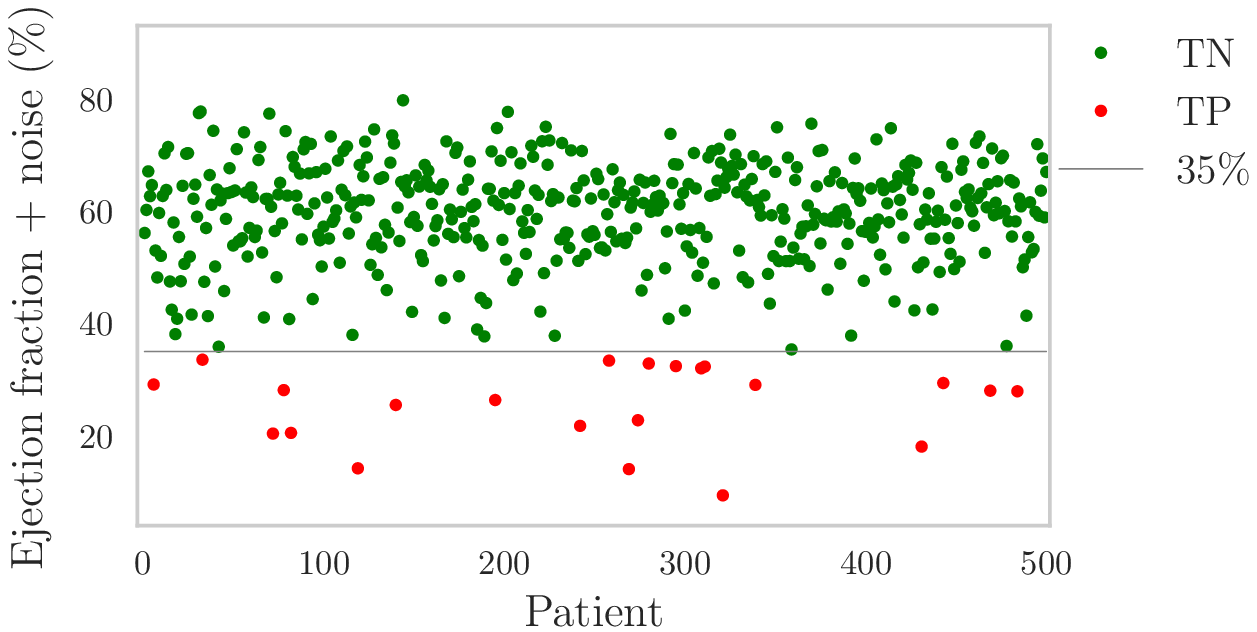}
		\caption{\Ac{EF} measurements.} \label{fig:ef1}
	\end{subfigure}\\
	\begin{subfigure}{.48\textwidth} \centering
		\includegraphics[height=4.3cm]{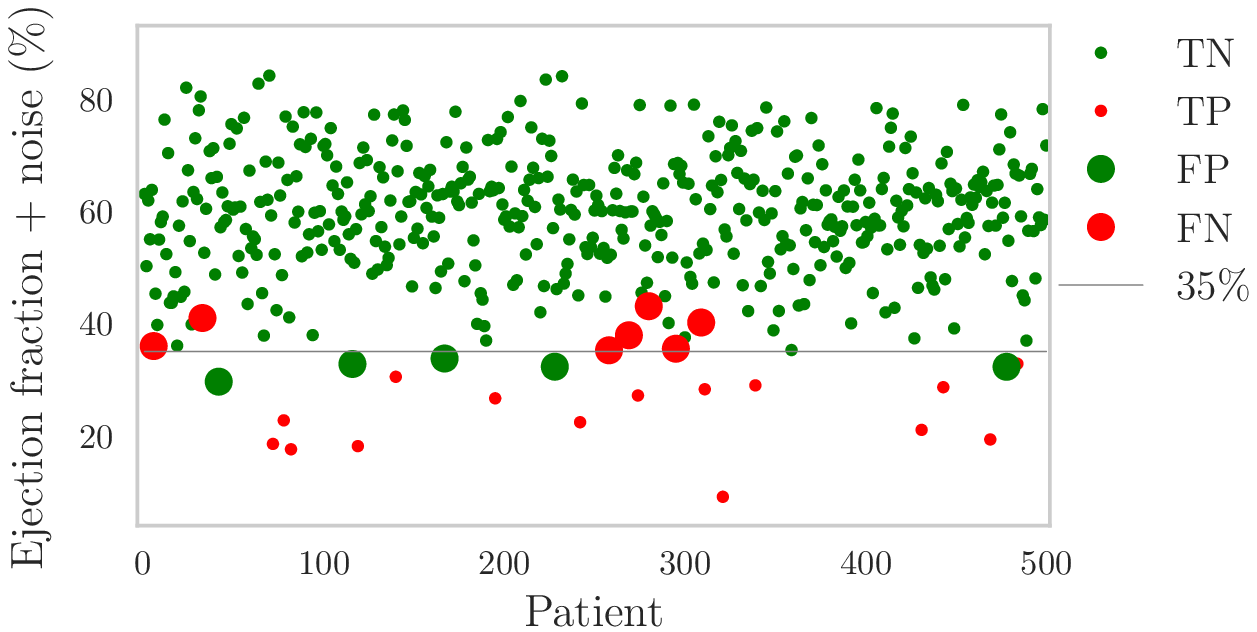}
		\caption{\Ac{EF} measurements with added noise.} \label{fig:ef2}
	\end{subfigure}
	\caption{ \footnotesize (a) \Acf{EF} data of the Data Science Bowl Cardiac Challenge. (b) Same data with noise sampled from $\mathcal{N}(0,0.1\bar{x})$, with $\bar{x}$ the mean. \Ac{EF} $< 35\%$ indicates therapy eligibility leading to \acf{TN}, \acf{TP}, \acf{FP}, \acf{FN}.} \label{fig:ef}
\end{figure}

We take the \acf{EF} estimates of the \emph{Data Science Bowl Cardiac Challenge}~\cite{kaggle2015}. \Ac{EF} is a variable of critical importance in cardiology. Implantable device therapy is officially recommended for $EF < 35\%$~\cite{Ponikowski2016}. Therefore, we take cases with $EF < 35\%$ to have positive eligibility, i.e. $Y= \mathbf{1}(EF < 35\%)$, as shown in Figure~\ref{fig:ef1}. Adding random noise to these data results in $7$ \acf{FN} and $5$ \acf{FP}, as seen in Figure~\ref{fig:ef2}.\\

\begin{figure}[ht]\centering
	\vspace{-3mm}
	\begin{subfigure}{.48\textwidth} \centering
		\includegraphics[trim={0cm 11.5cm 0cm 0cm}, clip, width=.95\textwidth]{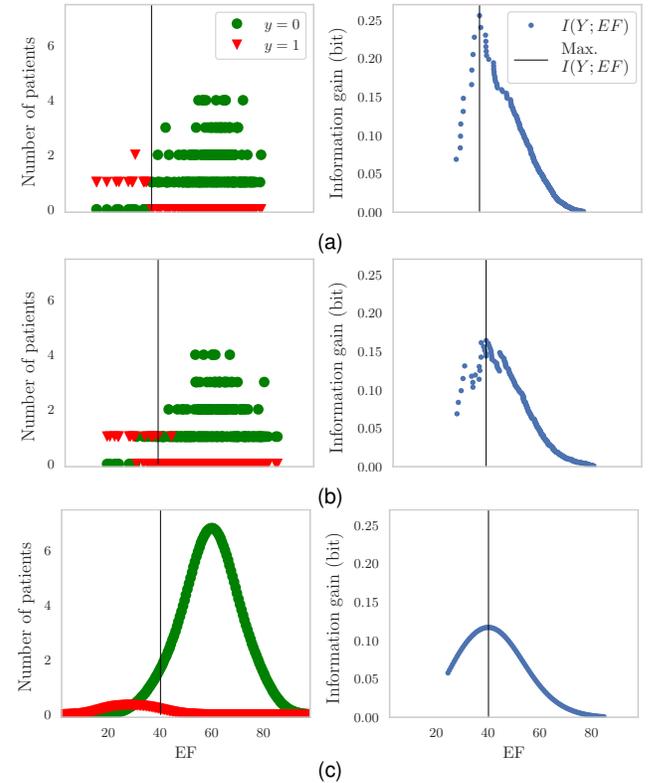} \caption{\scriptsize} \label{fig:ef-disc-a}
	\end{subfigure}
	\begin{subfigure}{.48\textwidth} \centering
		\includegraphics[trim={0cm 6.2cm 0cm 5.3cm}, clip, width=.95\textwidth]{figure3-2.eps} \caption{\scriptsize} \label{fig:ef-disc-b}
	\end{subfigure}
	\begin{subfigure}{.48\textwidth} \centering
		\includegraphics[trim={0cm .01cm 0cm 10.7cm}, clip, width=.95\textwidth]{figure3-2.eps} \vspace{-2mm} \caption{\scriptsize} \label{fig:ef-disc-c}
	\end{subfigure}
	\caption{ \footnotesize Information gain computation the \acf{EF} measurements of Figure~\ref{fig:ef}, using the standard search (a,b) or \acf{SS} with $u_s=0.1$ (c). The left (a) and (b) plots show the number of patients for each class and~$x^{(i)}$, $N(y=0, x^{(i)})$ and $N(y=1, x^{(i)})$. The left~(c) plot displays the \ac{SS} density increments, $\Delta\rho(y, \tau)$. The bottom plots show the corresponding information gain.}
	\label{fig:ef-disc}
\end{figure}

\noindent \textit{\Acl{SS}:} In Section~\ref{motivation}, we motivated the \ac{SS} as way of increasing the set of candidate splits and smoothing the information gain. We now observe this on real measurements. Figure~\ref{fig:ef-disc-a} displays the number of patients for each class and \ac{EF} value, $N(y, x^{(i)})$, like a histogram. Figure~\ref{fig:ef-disc-b} shows the same data with noise. Figure~\ref{fig:ef-disc-c} shows the \ac{SS} density increments $\Delta\rho(y, \tau)$, and \ac{SS} information gain.



\clearpage

Soft search methods such \ac{SS} or \ac{UDT} increase the set of candidate splits. This dataset has $500$ instances. Employing \ac{UDT} with a resampling factor $s=100$ raises the number of information gain computations from $5e2$ to $5e4$~\cite{Tsang2011}. Setting the \ac{SS} parameters e.g. as $\cal{w}=8$, $\delta=0.05$ and $u_s=0.1$ limits this number to a maximum of approximately~$2.3e3$.\\

\noindent \textit{\Acl{STP}:} To illustrate how \ac{STP} alters the class probability estimates, we consider the noisy data of Figure~\ref{fig:ef2}. We learned two single-node two-leaf trees using the standard training propagation and \ac{STP} with $u_{p} = 0.1$. Table~\ref{tab:tab1} shows that the class probability estimates are less extreme when \ac{STP} is employed, as they reflect the choice of noise distribution.\\

\begin{table}[t]
	\centering
	\caption{ Class probability estimates at leaves ${\mathcal{n}}_L$ and ${\mathcal{n}}_R$ of the trees learned on the noisy \acf{EF} dataset of Figure~\ref{fig:ef2}. \Acfp{DT} learned using standard or \acf{STP}.}
	\label{tab:tab1}
	\renewcommand{\arraystretch}{.9}
	\setlength{\tabcolsep}{6pt}
	\begin{tabular}{lcccc}
		\toprule
		& Standard training propagation & \Ac{STP} \\
		\midrule
		$\hat{p}(y=0|\mathcal{n}_L)$ & 0.000 & 0.310	\\
		$\hat{p}(y=1|\mathcal{n}_L)$ & 1.000 & 0.790	\\
		\midrule
		$\hat{p}(y=0|\mathcal{n}_R)$ & 0.981 & 0.979	\\
		$\hat{p}(y=1|\mathcal{n}_R)$ & 0.018 & 0.021	\\
		\bottomrule
	\end{tabular}
\end{table}
\begin{table}[t]
	\centering
	\caption{ Class probability estimates for the misclassified examples of the noisy \acf{EF} data in Figure~\ref{fig:ef2}, estimated by the tree learned with C4.5 on the non-noisy data in Figure~\ref{fig:ef1}.}
	\label{tab:stest}
	\renewcommand{\arraystretch}{.9}
	\setlength{\tabcolsep}{1.5pt}
	\begin{tabular}{llcccc}
		\toprule
		& & \multicolumn{2} {c} {Hard evaluation} & \multicolumn{2} {c}{\Ac{SE}}\\
		\midrule
		& Patient & $\hat{p}(y=0|\mathbf{x})$ & $\hat{p}(y=1|\mathbf{x})$ & $\hat{p}(y=0|\mathbf{x})$ & $\hat{p}(y=1|\mathbf{x})$ \\
		\midrule
		\multirow{6}{*}{False negative}		& 6		& 1.00 & 0.00 & 0.54		& 0.46 \\
		& 33		& 1.00 & 0.00 & 0.83		& 0.17 \\
		& 269	& 1.00 & 0.00 & 0.67		& 0.33 \\
		& 280	& 1.00 & 0.00 & 0.91		& 0.09 \\
		& 295	& 1.00 & 0.00 & 0.51		& 0.49 \\
		& 309	& 1.00 & 0.00 & 0.79		& 0.21 \\
		\midrule
		\multirow{6}{*}{False positive}		& 42		& 0.00 & 1.00 & 0.16		& 0.84 \\
		& 116	& 0.00 & 1.00 & 0.33		& 0.67 \\
		& 167	& 0.00 & 1.00 & 0.39		& 0.61 \\
		& 228	& 0.00 & 1.00 & 0.30		& 0.70 \\
		& 359	& 0.00 & 1.00 & 0.49		& 0.51 \\
		& 478	& 0.00 & 1.00 & 0.30		& 0.70 \\
		\bottomrule
	\end{tabular}
\end{table}

\noindent \textit{\Acl{SE}:} Figure~\ref{fig:ef2} shows that 7 \ac{FN} and 5 \ac{FP} were introduced by the noise added to the \ac{EF} dataset. Table~\ref{tab:stest} shows the probability estimates of those misclassified examples, obtained using hard or soft evaluation, with $u_e = 0.1$. The numbers of \ac{FN} and \ac{FP} are different because the algorithm learned a threshold of $35.37\%$ rather than $35\%$.

\subsection{Experimental results} \label{exp-res}

The average number of leaves, test accuracy and running time are computed over 30 train-test permutations, for each experiment and dataset. The absolute difference to C4.5, averaged over all datasets, is displayed in Tables~\ref{table:res-SS-prop-eval-train-noise} and~\ref{table:res-SS-prop-eval-test-noise}. 

\begin{figure*}
	\captionsetup[subfigure]{justification=centering} \centering
	\begin{subfigure}{\textwidth} \centering 
		\includegraphics[width=.8\textwidth]{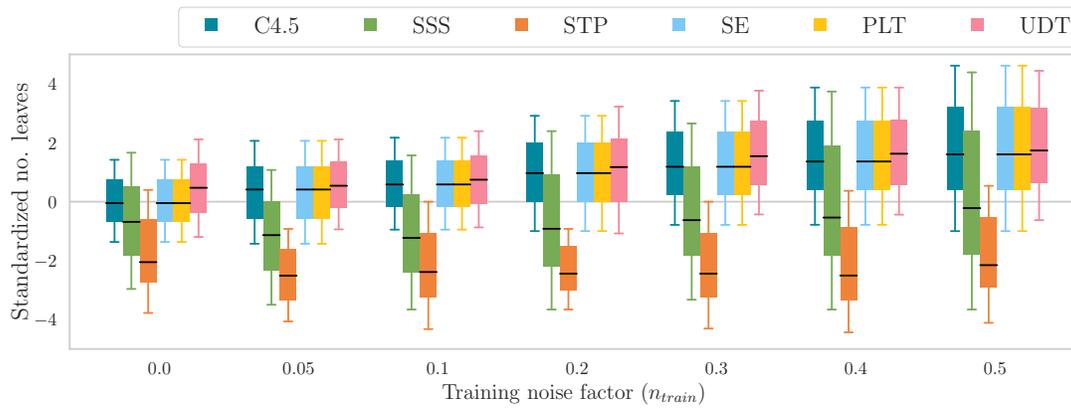} \vspace{-3mm}
		\caption{Standardized number of leaves of Experiment 1 - training data noise.\\~\\} \label{fig:res-train-noise-leaves}
	\end{subfigure}
	\begin{subfigure}{\textwidth} \centering 
		\includegraphics[width=.8\textwidth]{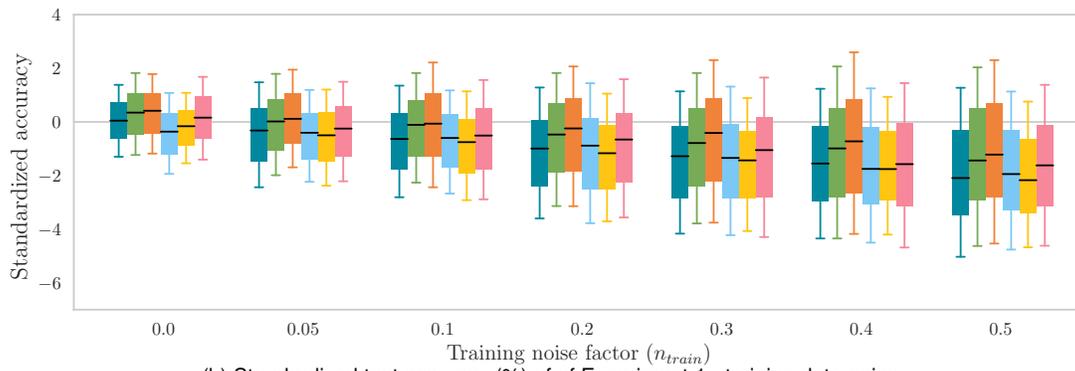} \vspace{-3mm}
		\vspace{-1mm}
		\caption{Standardized test accuracy (\%) of of Experiment 1 - training data noise.\\~\\} \label{fig:res-train-noise-acc}
	\end{subfigure}
	
	\begin{subfigure}{\textwidth} \centering 
		\includegraphics[width=.8\textwidth]{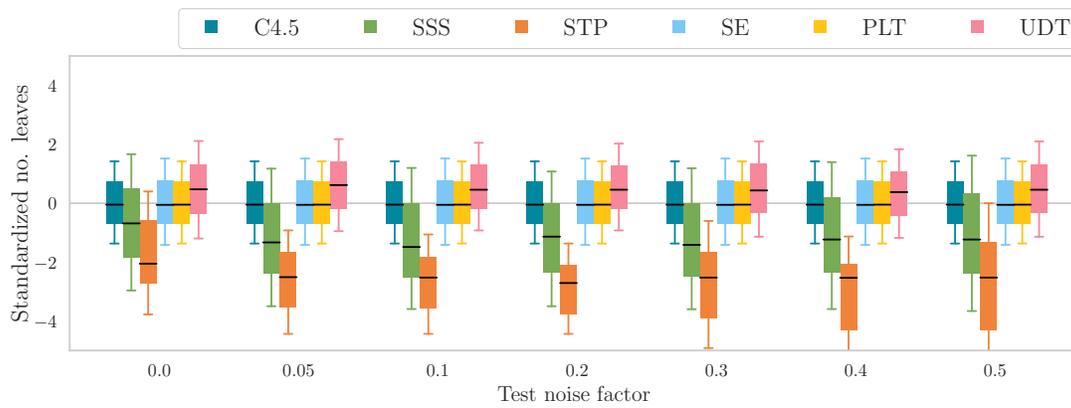} \vspace{-3mm}
		
		\caption{Standardized number of leaves of Experiment 2 - test data noise.\\~\\} \label{fig:res-test-noise-leaves}
	\end{subfigure}
	\begin{subfigure}{\textwidth} \centering 
		\includegraphics[width=.8\textwidth]{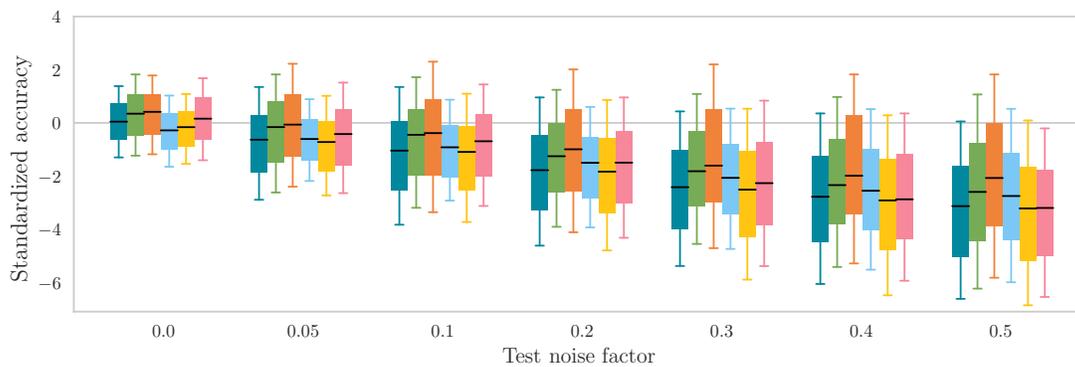} \vspace{-3mm}
		\caption{Standardized test accuracy (\%) of Experiment 2 - test data noise.\\~\\} \label{fig:res-test-noise-acc}
	\end{subfigure}
	
	\caption{Results of the experiments displayed as boxplots of the standardized metrics for all datasets. (a,b)~-~Experiment 1: models trained on data with increasing levels of  noise~$\sim~\mathcal{N}(0, n_{train}\bar{x})$, and evaluated on data without added noise. (c,d)~-~Experiment 2: models trained on data without added noise, and evaluated on data with noise~$\sim~\mathcal{N}(0, n_{test}\bar{x})$. The baseline was obtained with~C4.5.}
	\label{fig:res-box-plots}
\end{figure*}

However, since absolute results of distinct datasets cannot be directly compared~\cite{Demsar2006}, we focus on standardized metrics. The results of each dataset and method were standardized by the dataset's baseline. The baseline result is obtained by C4.5 without added noise, and estimated with the 30 permutations. The standardization consists in subtracting the baseline mean, and dividing by the baseline standard deviation. E.g. the baseline of the Heart disease dataset has mean $15.0$ leaves and standard deviation $3.0$ leaves. In this case, standardized results $0.0$, $-1.0$ and $1.5$ translate into $15.0$, $12.0$ and $19.5$ leaves, respectively. Tables~\ref{table:res-SS-prop-eval-train-noise} and~\ref{table:res-SS-prop-eval-test-noise} display the standardized metrics, averaged over all datasets, and Figure~\ref{fig:res-box-plots} shows the corresponding boxplots. Computational times are merely indicative, as the experiments were run on a cluster, and the specifications of the machines may vary slightly. 

All approaches show an increase in the number of leaves and a reduction in test accuracy as the noise factor grows. The decrease in accuracy was sharper for the predictions made on noisy test data, as seen in Table~\ref{table:res-SS-prop-eval-test-noise}.

\Ac{SS}, \ac{STP} and \ac{UDT} show maintenance or non-statistically significant improvement in accuracy compared to C4.5, in all noise scenarios. In Table~\ref{table:res-SS-prop-eval-train-noise}, we see that \ac{STP} had higher accuracy compared to the other methods for all $n_{train}$ and~$n_{test}$. For the \ac{SS} and \ac{STP} approaches, the maintenance of accuracy was accompanied by statistically significant reductions in the number of leaves compared to C4.5 and \ac{UDT}. The \ac{SS} tree size reduction was statistically significant for noise factors greater than $0.00$. \Ac{STP} had a further reduction in tree size, significant for all $n_{train}$ and~$n_{test}$. The maintenance of accuracy by \ac{UDT} compared to C4.5 was accompanied by an increase in the number of leaves. The method was considerably slower than \ac{SS} and~\ac{STP}.

In Experiment~1, the accuracies obtained by \ac{SE} and \ac{PLT} were equal or smaller than those obtained through hard evaluation, as seen in Table~\ref{table:res-SS-prop-eval-train-noise}. This is particularly evident when~$n_{train} = 0$. The number of leaves remains unchanged as these methods do not affect training.

On the contrary, Table~\ref{table:res-SS-prop-eval-test-noise} shows that \ac{SE} accuracy was superior to that of hard evaluation in Experiment~2 for $n_{test} \ge 0.05$. However, this increase was not statistically significant. It suggests that the uncertainty distribution considered in the \ac{SE} approach better captures the noise added to the data, compared to~\ac{PLT}.

\section{Discussion}

We propose a probabilistic \ac{DT} approach to handle uncertain data, which separates the uncertainty model in three independent algorithm components. Our experiments evaluate these components in their ability to handle varying degrees of noise in the training and test data.

The first observation is that corrupting the data decreases the accuracy of the predictions, specially if the noise is in the test data. Accordingly, learning on data with increasing uncertainty results in \acp{DT} with a larger number of leaves, as the models attempt to learn the particularities of the training set.	

The results indicate that \ac{SS}, \ac{STP} and \ac{UDT} are at least as robust to noise as C4.5, with non-significant improvements in accuracy. This was observed both for training or test data noise. \Ac{UDT} results are consistent with previous experiments with accuracy gains in the order of $3\%$, but where \ac{DT} size was not reported~\cite{Tsang2011}.

All soft training methods had longer running times compared to C4.5, the slowest being~\ac{UDT}. When comparing the search approaches, we observe that, by employing the discretization in Equation~\ref{eqn:discretization}, \ac{SS} increases the set of possible thresholds compared to C4.5, while preventing the computation of the information gain for values $x^{(i)}$ that are very close. \Ac{UDT} generates $s$ samples for each measurement. The number of entropy computations per attribute is bounded by~$s |\mathcal{D}|$, and therefore grows with the size of the dataset. Using \ac{SS}, this number is bounded by $\tfrac{1}{\delta} ( \tau_{max} - \tau_{min} )$, and does not grow with $|\mathcal{D}|$.

While maintaining accuracy, \ac{SS} and \ac{STP} led to significantly smaller~\acp{DT}. All approaches built larger trees for increasing $n_{train}$, as they start to overfit to the noise. \Ac{SS} and \ac{STP} were able to cope with this by reducing the number of splits, acting as regularizers. This can be interpreted as a consequence of having class probability estimates that reflect the uncertainty, illustrated in Figure~\ref{fig:perror-ex1a} and Table~\ref{tab:tab1}, on the pruning algorithm. The C4.5 pruning method uses a statistical test on the training data to make a pessimistic estimate the generalization error. Using a soft training approach expresses less confidence in the data, causing the pruning algorithm to remove more nodes. Tree size reductions were also observed for the multivariate sigmoid-split approach~\cite{irsoy2012soft}. However, they were most likely caused by the use of multivariate split functions, which are able to express complex rules more compactly, at the cost of reduced interpretability.


To investigate if the \ac{DT} size reduction could be obtained by changing the C4.5 pruning confidence factor $c$, in Appendix~C we show the result of varying $c$ when using C4.5, \ac{SS} or \ac{STP} for the first $5$ datasets of Table~\ref{tab:datasets}. For lower $c$, \ac{SS} resulted in smaller models with similar accuracy. In the overfitting range, this tendency is inverted. This indicates that the estimated class probabilities are more accurate with \ac{SS}, when the model actually learned representative splits. \Acp{STP} has led to consistently smaller trees than C4.5, except when $c$ is vey close to~$1$.

Given that the uncertainty models in \ac{SS}, \ac{STP} and \ac{UDT} are all Gaussian, an explanation for the disparity in results obtained by the proposed soft training approaches and \ac{UDT} could be the mismatch between the uncertainty model considered in the algorithms and the distribution of the noise added to the data. In our experiments, the same distribution was used in \ac{SS} and \ac{STP} and in the noise model, i.e. the standard deviation was a factor of the variable mean. In~\ac{UDT}, the standard deviation is instead a factor of the range of the data. This is also suggested by the values of the $w$ parameter selected through \ac{CV} displayed in Figures \ref{fig:params-train-noise} and~\ref{fig:params-test-noise}, which are smaller than $u_s$ and~$u_t$. Some degree of correlation between the parameters $u_s$, $u_t$ and the noise factors $n_{train}$, $n_{test}$ was observed. But no significant correlation was observed between $w$ and $n_{train}$ or $n_{test}$.

We hypothesize that the soft learning works more effectively as the uncertainty model approximates the real noise distribution. We therefore recommend the use of uncertainty models specified by domain experts. In our experiments, noise was simulated as an experimental proof-of-concept. To validate our approach on concrete clinical decision problems, the uncertainty distribution and its parameters shall be estimated \textit{a priori} for each variable. Such an estimation may be based, for instance, on the meta-analysis of clinical studies and on empirical clinical knowledge.

\begin{table}[ht]
	\centering
	\setlength{\tabcolsep}{4pt}
	\caption[]{{ \footnotesize Results for Experiment~1. Standardized metrics and absolute difference to C4.5 are presented, averaged over all datasets. The standardized results are tested for statistically significant differences according to a Wilcoxon signed-ranks test~\cite{Wilcoxon1945}, where each result is tested against C4.5 with the same~$n_{train}$. Standardized results with $p<.001$ highlighted in bold.}}
	\scriptsize
	\begin{tabular}{c|l|rrrrrrr}
		\toprule
		\multirow{2}{*}{Metric} 	& \multirow{2}{*}{Method}	& \multicolumn{7}{c}{Training data noise factor ($n_{train}$)} \\
		& 							& 0.00 & 0.05 & 0.10 & 0.20 & 0.30 & 0.40 & 0.50\\
		\midrule \midrule
		\multirow{6}{*}{\specialcell{No. leaves \\ standardized \\ by  C4.5 with \\ $n_{train}=0$}} 
		& C4.5		& 0.00	& 0.44	& 0.72	& 1.09	& 1.42	& 1.75	& 2.03	\\
		& \Ac{SS}		& -0.48			 &	\textbf{-0.78} &	\textbf{-0.81} &	\textbf{-0.41} &	\textbf{-0.01} &	\textbf{0.40} &	\textbf{0.75} \\
		& \Ac{STP}		& \textbf{-1.11} &	\textbf{-2.19} &	\textbf{-2.06} &	\textbf{-1.95} &	\textbf{-0.90} &	\textbf{-1.01}	&	\textbf{-0.16} \\
		& \Ac{SE}		& 0.00	& 0.44	& 0.72	& 1.09	& 1.42	& 1.75	& 2.03	\\
		& \Ac{PLT}		& 0.00	& 0.44	& 0.72	& 1.09	& 1.42	& 1.75	& 2.03	\\
		& \acs{UDT}		& 0.46	& 0.64	& 0.83	& 1.25	& 1.81	& 1.97	& 2.17	\\
		\midrule
		
		\multirow{5}{*}{\specialcell{Absolute diff. \\ in no. leaves \\ to C4. with \\ same $n_{train}$ }}
		& \Ac{SS}		& -2.9	& -5.1 	& -5.6	& -5.4	& -5.4	& -5.3	& -5.0	\\
		& \Ac{STP}		& -5.4 	& -8.4 	& -8.6 	& -9.3	& -8.9	& -9.3	& -8.7	\\
		& \Ac{SE}		& 0.0	& 0.0	& 0.0	& 0.0	& 0.0	& 0.0	& 0.0	\\
		& \Ac{PLT}		& 0.0	& 0.0	& 0.0	& 0.0	& 0.0	& 0.0	& 0.0	\\
		& \acs{UDT}		& 1.5	& 0.43	& 0.6	& 0.7	& 1.1	& 0.4	& 0.2	\\
		\midrule
		\midrule
		
		\multirow{6}{*}{\specialcell{Accuracy \\ standardized \\ by C4.5 with \\ $n_{train}=0$}}
		& C4.5		& 0.00		& -0.52		& -0.87		& -1.27		& -1.59		& -1.65		& -2.06		\\
		& \Ac{SS}	& 0.26		& -0.14		& -0.39		& -0.79		& -1.05		& -1.23		& -1.40 	\\
		& \Ac{STP}	& 0.32		& -0.03		& -0.30		& -0.70		& -0.86		& -1.06		& -1.24		\\
		& \Ac{SE}	& -0.60		& -0.57		& -0.87		& -1.27		& -1.59		& -1.81		& -1.94		\\
		& \Ac{PLT}	& -0.51		& -0.67		& -1.01		& -1.38		& -1.68		& -1.75		& -2.14 	\\
		& \acs{UDT}	& 0.09		& -0.45		& -0.72		& -1.10		& -1.36		& -1.70		& -1.81 	\\
		\midrule
		
		\multirow{5}{*}{\specialcell{Absolute diff. \\ in accuracy (\%) \\ to C4. with \\ same $n_{train}$ }}
		& \Ac{SS}	& 0.8		& 1.2		& 1.7		& 1.6		& 1.8	&	1.3		&	2.1		\\
		& \Ac{STP}	& 0.9		& 1.6		& 2.0		& 1.9		& 2.3	&	1.9		&	2.6		\\
		& \Ac{SE}	& -2.0		& -0.3		& 0.0		& -0.1		& 0.0	&	-0.5	&	0.3		\\
		& \Ac{PLT}	& -1.2		& -0.5		& -0.4		& -0.3		& -0.2	&	-0.2	&	-0.2	\\
		& \acs{UDT}	& 0.3		& 0.1		& 0.5		& 0.4		& 0.6	&	-0.3	&	0.7		\\
		\midrule \midrule
		
		\multirow{5}{*}{\specialcell{Absolute diff. \\ running time (s) \\ to C4. with \\ same $n_{train}$ }}
		& \Ac{SS}	& 2.7	& 5.6	& 4.5	& 4.9	& 5.1	& 4.3	& 5.2	\\
		& \Ac{STP}	& 1.0	& 1.2	& 1.4	& 1.9	& 1.9	& 1.5	& 1.5	\\
		& \Ac{SE}	& 0.0	& 0.1	& -0.1	& 0.1	& 0.0	& 0.1	& 0.2	\\
		& \Ac{PLT}	& 0.1	& 0.2	& 0.2	& 0.2	& 0.2	& 0.2	& 0.2	\\
		& \acs{UDT}	& 40.6	& 45.0	& 37.6	& 49.0	& 53.9	& 37.3	& 44.2	\\
		
		\bottomrule
	\end{tabular} \label{table:res-SS-prop-eval-train-noise}
\end{table}
\begin{table}[ht]
	\centering
	\setlength{\tabcolsep}{4.2pt}
	\caption[]{{ \footnotesize Results for Experiment~2. Standardized metrics and absolute difference to C4.5 are presented, averaged over all datasets. The standardized results are tested for statistically significant differences according to a Wilcoxon signed-ranks test~\cite{Wilcoxon1945}, where each result is tested against C4.5 with the same~$n_{test}$. Standardized results with $p<.001$ highlighted in bold.}}
	\scriptsize
	\begin{tabular}{l|l|rrrrrrr}
		\toprule
		\multirow{2}{*}{Metric} 	& \multirow{2}{*}{Method}	& \multicolumn{7}{c}{Test data noise factor ($n_{test}$)} \\
		& 							& 0.00 & 0.05 & 0.10 & 0.20 & 0.30 & 0.40 & 0.50\\ 
		\midrule \midrule
		\multirow{6}{*}{\specialcell{No. leaves \\ standardized \\ by  C4.5 with \\ $n_{test}=0$}} 
		& C4.5			& 0.00	& 0.00	& 0.00	& 0.00	& 0.00	& 0.00	& 0.00	\\
		&	\Ac{SS}	& -0.48				&	\textbf{-1.10}	& \textbf{-1.07}	& \textbf{-0.81}	& \textbf{-0.93}	& \textbf{-0.81}	& \textbf{-0.77}	\\
		&	\Ac{STP}	& \textbf{-1.11}	&	\textbf{-1.73}	& \textbf{-1.89}	& \textbf{-2.05}	& \textbf{-1.90}	& \textbf{-2.16}	& \textbf{-1.98}	\\
		&	\Ac{SE}		& 0.00				&	0.00			& 0.00				& 0.00				& 0.00				& 0.00				& 0.00				\\
		&	\Ac{PLT}	& 0.00				&	0.00			& 0.00				& 0.00				& 0.00				& 0.00				& 0.00				\\
		&	\acs{UDT}	& 0.46				&	0.62			& 0.49				& 0.45				& 0.47				& 0.33				& 0.49				\\
		\midrule
		
		\multirow{5}{*}{\specialcell{Absolute diff. \\ in no. leaves \\ to C4. with \\ same $n_{test}$ }}
		& \Ac{SS}		& -2.9	& -4.3	& -4.3	& -3.4	& -4.0	& -3.5	& -3.4	\\
		& \Ac{STP}		& -5.4	& -7.0	& -7.3	& -7.7	& -7.5	& -7.7	& -7.5	\\
		& \Ac{SE}		& 0.0	& 0.0	& 0.0	& 0.0	& 0.0	& 0.0	& 0.0	\\
		& \Ac{PLT}		& 0.0	& 0.0	& 0.0	& 0.0	& 0.0	& 0.0	& 0.0	\\
		& \acs{UDT}		& 1.5	& 1.7	& 1.3	& 1.2	& 0.9	& 0.9	& 1.5	\\
		\midrule \midrule
		
		\multirow{6}{*}{\specialcell{Accuracy \\ standardized \\ by C4.5 with \\ $n_{test}=0$}}
		& C4.5								& 0.00				& -0.97				& -1.38				& -2.01		& -2.59	& -3.06	& -3.50	\\
		& \Ac{SS}							& 0.26				& -0.50				& -0.82				& -1.45		& -1.85 & -2.27	& -2.62	\\
		& \Ac{STP}							& 0.32				& -0.40				& -0.72				& -1.18		& -1.54	& -1.91	& -2.15	\\
		& \Ac{SE}							& -0.60				& -0.85				& -1.19				& -1.79		& -2.32	& -2.78	& -3.19	\\
		& \Ac{PLT}							& -0.51				& -1.16				& -1.60				& -2.28		& -2.90	& -3.43	& -3.86 \\
		& \acs{UDT}							& 0.09				& -0.64				& -1.02				& -1.74		& -2.39	& -2.94	& -3.46 \\
		\midrule
		
		\multirow{5}{*}{\specialcell{Absolute diff. \\ in accuracy (\%) \\ to C4. with \\ same $n_{test}$ }}
		& \Ac{SS}	& 0.8	&	1.5		&	2.0		&	1.8		&	2.2		&	2.4		&	2.7		\\
		& \Ac{STP}	& 0.9	&	1.7		&	2.1		&	2.5		&	3.2		&	3.4		&	4.1		\\
		& \Ac{SE}	& -2.0	&	-0.2	&	0.0	&	-0.1	&	-0.1		&	-0.2		&	-0.1		\\
		& \Ac{PLT}	& -1.2	&	-0.5	&	-0.5	&	-0.6	&	-0.6	&	-0.7	&	-0.7	\\
		& \acs{UDT}	& 0.3	&	0.6		&	0.8		&	0.5		&	0.3		&	0.1		&	-0.1	\\
		\midrule \midrule
		
		\multirow{5}{*}{\specialcell{Absolute diff. \\ running time (s) \\ to C4. with \\ same $n_{train}$ }}
		& \Ac{SS}	& 2.05	& 4.8	& 6.0	& 4.8	& 4.7	& 3.3	& 5.29	\\
		& \Ac{STP}	& 1.7	& 2.8	& 2.8	& 3.2	& 3.7	& 3.4	& 3.40	\\
		& \Ac{SE}	& 0.1	& 0.1	& 0.1	& 0.1	& 0.0	& 0.0	& 0.0	\\
		& \Ac{PLT}	& 0.0	& 0.0	& 0.0	& 0.0	& 0.0	& 0.0	& 0.0	\\
		& \acs{UDT}	& 40.9	& 37.7	& 33.9	& 38.1	& 39.0	& 40.3	& 39.9	\\
		
		\bottomrule
		
	\end{tabular} \label{table:res-SS-prop-eval-test-noise}
\end{table}

\clearpage


\clearpage

\Ac{SE} and \ac{PLT} did not improve accuracy given noisy training data, compared to the standard hard split approach. When noise was added to the test data, \ac{SE} led to non-significant increases in accuracy. This suggests that modeling uncertainty to target training data noise is only effective when this model is incorporated in the training phase, and not during evaluation. As such, we do not recommend the use of soft evaluation to handle training noise.

The disparity between the \ac{SE} and the \ac{PLT} results may also be explained by the consistency between the uncertainty model considered by the algorithms, and the model used to corrupt the data. \Ac{PLT} has demonstrated 2-3\% error rate reductions on a previous experiment using a single dataset, where the shape of the uncertainty model had been optimized.

\section{Conclusions}

This paper presents a probabilistic \ac{DT} learning approach to handle the uncertainty in the data, where the uncertainty model is separated in three independent algorithm components. The context is providing interpretable models for clinical decision support, with the motivation that the acknowledgement of uncertainty will facilitate the adoption of automated learning approaches in practice. 

Previous \ac{DT} algorithms suggest the potential of probabilistic approaches to improve prediction robustness, and the need for an evaluation on more datasets and levels of noise. The impact of considering an uncertainty model in the learning phase or during evaluation was not however reported, as well as the impact of having noise is the training examples or the test examples.

In our approach, the uncertainty representation is incorporated: in the learning phase when searching for the optimal thresholds (\ac{SS}), when propagating the training data through the tree (\ac{STP}), and in the evaluation phase when obtaining predictions for unseen data (\ac{SE}). Any model can be chosen to capture the uncertainty. Our purpose is to incorporate clinical knowledge about the reliability of measurements. But as a proof-of-concept, we model the uncertainty as normally-distributed noise.

In our experiments, corrupting the test data seems to have a more severe impact on accuracy than adding noise to the training data. Upon increased noise, the soft training components, \ac{SS}, \ac{STP} and \ac{UDT}, show maintained or improved accuracy compared to C4.5. \Ac{STP}~and \ac{SS} act as regularizers, showing significant reductions in tree size, with \ac{STP} outperforming the latter. This was not the case of \ac{UDT}, possibly given the disparity between the noise model in the data and the uncertainty model in the algorithm. The running times of \ac{SS} and \ac{STP} were lower than those of \ac{UDT}. None of the soft evaluation approaches shows significant benefit compared to hard evaluation. Overall, we recommend using \ac{SS} and \ac{STP} with an uncertainty model that approximates as much as possible the real noise in the data. Finally, our study shows the importance of the acknowledgement of data uncertainty when learning decision models. Ideally, when designing clinical studies, an assessment of the reliability of each measurement should be considered part of the database.

Future work directions include evaluating the approach with domain-specific noise distributions, and studying the conditions under which the soft training provides benefit, regarding the complexity of the data. For highly separable data, the benefit of any soft approach is expected to be limited.

\ifCLASSOPTIONcompsoc
  \section*{Acknowledgments}
 This work was supported by the European Union Horizon 2020 research and innovation programme under grant agreement 642676 (Cardiofunxion), by the Spanish Ministry of Economy and Competitiveness (grant TIN2014-52923-R; Maria de Maeztu Units of Excellence Programme - MDM-2015-0502), by the European Union FP7 for research, technological development and demonstration under grant agreement VP2HF (611823), and FEDER.\\
\else
  \section*{Acknowledgment}
\fi

\ifCLASSOPTIONcaptionsoff
  \newpage
\fi

\IEEEtriggeratref{32}

\bibliographystyle{IEEEtran}
\bibliography{bib}

\begin{thebibliography}{10}
\providecommand{\url}[1]{#1}
\csname url@samestyle\endcsname
\providecommand{\newblock}{\relax}
\providecommand{\bibinfo}[2]{#2}
\providecommand{\BIBentrySTDinterwordspacing}{\spaceskip=0pt\relax}
\providecommand{\BIBentryALTinterwordstretchfactor}{4}
\providecommand{\BIBentryALTinterwordspacing}{\spaceskip=\fontdimen2\font plus
\BIBentryALTinterwordstretchfactor\fontdimen3\font minus
  \fontdimen4\font\relax}
\providecommand{\BIBforeignlanguage}[2]{{%
\expandafter\ifx\csname l@#1\endcsname\relax
\typeout{** WARNING: IEEEtran.bst: No hyphenation pattern has been}%
\typeout{** loaded for the language `#1'. Using the pattern for}%
\typeout{** the default language instead.}%
\else
\language=\csname l@#1\endcsname
\fi
#2}}
\providecommand{\BIBdecl}{\relax}
\BIBdecl

\bibitem{Patel2009}
V.~L. Patel, E.~H. Shortliffe, M.~Stefanelli, P.~Szolovits, M.~R. Berthold,
  R.~Bellazzi, and A.~Abu-Hanna, ``{The coming of age of artificial
  intelligence in medicine},'' \emph{Artificial Intelligence in Medicine},
  vol.~46, no.~1, pp. 5--17, 2009.

\bibitem{Clifton2015}
D.~A. Clifton, K.~E. Niehaus, P.~Charlton, and G.~W. Colopy, ``{Health
  Informatics via Machine Learning for the Clinical Management of Patients},''
  \emph{Yearb Med Inform}, vol.~10, no.~1, pp. 38--43, 2015.

\bibitem{Roddick2003}
J.~Roddick, P.~Fule, and W.~Graco, ``{Exploratory medical knowledge discovery:
  Experiences and issues},'' \emph{ACM SIGKDD Explorations Newsletter}, pp.
  2--7, 2003.

\bibitem{Lavrac1998}
N.~Lavra{\v{c}}, I.~Kononenko, E.~Keravnou, M.~Kukar, and B.~Zupan,
  ``{Intelligent data analysis for medical diagnosis: Using machine learning
  and temporal abstraction},'' \emph{AI Communications}, vol.~11, no.~3, pp.
  191--218, 1998.

\bibitem{Karr2006}
A.~F. Karr, A.~P. Sanil, and D.~L. Banks, ``{Data quality: A statistical
  perspective},'' \emph{Statistical Methodology}, vol.~3, no.~2, pp. 137--173,
  2006.

\bibitem{jcgm2008evaluation}
W.~G.~. Joint Committee for Guides~in Metrology, ``Evaluation of measurement
  data - guide to the expression of uncertainty in measurement,'' in
  \emph{Tech. Rep. JCGM 100: 2008 (BIPM, IEC, IFCC, ILAC, ISO, IUPAC, IUPAP and
  OIML}, 2008.

\bibitem{Cios2002}
K.~J. Cios and G.~{William Moore}, ``{Uniqueness of medical data mining},''
  \emph{Artificial Intelligence in Medicine}, vol.~26, no. 1-2, pp. 1--24,
  2002.

\bibitem{Singh2003}
K.~Singh, B.~K. Jacobsen, S.~Solberg, K.~H. B{\o}naa, S.~Kumar, R.~Bajic, and
  E.~Arnesen, ``{Intra- and interobserver variability in the measurements of
  abdominal aortic and common iliac artery diameter with computed tomography.
  The Troms{\o} study},'' \emph{European Journal Vascular and Endovascular
  Surgery}, vol.~25, no.~5, pp. 399--407, 2003.

\bibitem{Foley2012}
T.~Foley, S.~Mankad, N.~Anavekar, C.~Bonnichsen, M.~Morris, T.~Miller, and
  P.~Araoz, ``{Measuring left ventricular ejection fraction-techniques and
  potential pitfalls},'' \emph{European Cardiology}, vol.~8, no.~2, pp.
  108--114, 2012.

\bibitem{Lopez-Minguez2014}
J.~R. Lopez-Minguez, R.~Gonzalez-Fernandez, C.~Fernandez-Vegas,
  V.~Millan-Nunez, M.~E. Fuentes-Canamero, J.~M. Nogales-Asensio,
  J.~Doncel-Vecino, M.~{Yuste Dominguez}, L.~{Garcia Serrano}, and D.~{Sanchez
  Quintana}, ``{Comparison of imaging techniques to assess appendage anatomy
  and measurements for left atrial appendage closure device selection.}''
  \emph{The Journal of invasive cardiology}, vol.~26, no.~9, pp. 462--467, sep
  2014.

\bibitem{genders2017quantitative}
T.~S. Genders, B.~S. Ferket, and M.~M. Hunink, ``The quantitative science of
  evaluating imaging evidence,'' \emph{JACC: Cardiovascular Imaging}, vol.~10,
  no.~3, pp. 264--275, 2017.

\bibitem{DeHaan2014}
S.~de~Haan, K.~de~Boer, J.~Commandeur, A.~M. Beek, A.~C. van Rossum, and C.~P.
  Allaart, ``{Assessment of left ventricular ejection fraction in patients
  eligible for ICD therapy: Discrepancy between cardiac magnetic resonance
  imaging and 2D echocardiography},'' \emph{Netherlands Heart Journal},
  vol.~22, no.~10, pp. 449--455, 2014.

\bibitem{green2014closing}
L.~W. Green, ``Closing the chasm between research and practice: evidence of and
  for change,'' \emph{Health Promotion Journal of Australia}, vol.~25, no.~1,
  pp. 25--29, 2014.

\bibitem{quinlan1979interactive}
J.~Quinlan \emph{et~al.}, ``Interactive dichotomizer, id3,'' \emph{Eds. Morgan
  Kauffmann, Springer-Verlag}, 1979.

\bibitem{Quinlan1993}
R.~Quinlan, \emph{C4.5: Programs for Machine Learning}.\hskip 1em plus 0.5em
  minus 0.4em\relax San Mateo, CA: Morgan Kaufmann Publishers, 1993.

\bibitem{Breiman1984}
L.~Breiman, J.~H. Friedman, R.~A. Olshen, and C.~J. Stone, \emph{Classification
  and Regression Trees}.\hskip 1em plus 0.5em minus 0.4em\relax Belmont, CA:
  Wadsworth International Group, 1984.

\bibitem{kass1980exploratory}
G.~V. Kass, ``An exploratory technique for investigating large quantities of
  categorical data,'' \emph{Applied statistics}, pp. 119--127, 1980.

\bibitem{Weng2017}
S.~F. Weng, J.~Reps, J.~Kai, J.~M. Garibaldi, and N.~Qureshi, ``Can
  machine-learning improve cardiovascular risk prediction using routine
  clinical data?'' \emph{{PLOS} {ONE}}, vol.~12, no.~4, 2017.

\bibitem{EU}
``{Regulation (EU) 2016/679 of the European Parliament and of the Council of 27
  April 2016 on the protection of natural persons with regard to the processing
  of personal data and on the free movement of such data, and repealing
  Directive 95/46/EC},'' 2016 O.J. L 119, 4.5.:1–88.

\bibitem{Quinlan1987}
J.~R. Quinlan, ``Decision trees as probabilistic classifiers,'' in
  \emph{Proceedings of the 4th International Workshop on Machine
  Learning}.\hskip 1em plus 0.5em minus 0.4em\relax Morgan Kauffman, 1987, pp.
  31--37.

\bibitem{DvorakS07}
J.~Dvor{\'{a}}k and P.~Savick{\'{y}}, ``Softening splits in decision trees
  using simulated annealing,'' in \emph{Adaptive and Natural Computing
  Algorithms, 8th International Conference, {ICANNGA} 2007, Warsaw, Poland,
  April 11-14, 2007, Proceedings, Part {I}}, 2007, pp. 721--729.

\bibitem{Tsang2011}
S.~Tsang, B.~Kao, K.~Y. Yip, W.-S. Ho, and S.~D. Lee, ``Decision trees for
  uncertain data,'' \emph{IEEE transactions on knowledge and data engineering},
  vol.~23, no.~1, pp. 64--78, 2011.

\bibitem{irsoy2012soft}
O.~Irsoy, O.~T. Y{\i}ld{\i}z, and E.~Alpayd{\i}n, ``Soft decision trees,'' in
  \emph{Pattern Recognition (ICPR), 2012 21st International Conference
  on}.\hskip 1em plus 0.5em minus 0.4em\relax IEEE, 2012, pp. 1819--1822.

\bibitem{Yuan1995}
Y.~Yuan, ``{Induction of fuzzy decision trees},'' \emph{Fuzzy Sets and
  Systems}, vol.~69, no.~2, pp. 125--139, 1995.

\bibitem{Wang2000}
X.~Wang, B.~Chen, G.~Qian, and F.~Ye, ``{On the optimization of fuzzy decision
  trees},'' \emph{Fuzzy Sets and Systems}, vol. 112, no.~1, pp. 117--125, may
  2000.

\bibitem{Segatori2017}
A.~Segatori, F.~Marcelloni, and W.~Pedrycz, ``{On Distributed Fuzzy Decision
  Trees for Big Data},'' \emph{IEEE Transactions on Fuzzy Systems}, pp. 1--1,
  2017.

\bibitem{quinlan1990probabilistic}
J.~R. Quinlan, ``Probabilistic decision trees,'' \emph{Machine learning: an
  artificial intelligence approach}, vol.~3, pp. 140--152, 1990.

\bibitem{jordan1994hierarchical}
M.~I. Jordan and R.~A. Jacobs, ``Hierarchical mixtures of experts and the em
  algorithm,'' \emph{Neural computation}, vol.~6, no.~2, pp. 181--214, 1994.

\bibitem{Hyafil1976}
L.~Hyafil and R.~L. Rivest, ``{Constructing optimal binary decision trees is
  NP-complete},'' \emph{Information Processing Letters}, vol.~5, no.~1, pp.
  15--17, 1976.

\bibitem{Quinlan1986}
J.~R. Quinlan, ``{Induction of decision trees},'' \emph{Machine Learning},
  vol.~1, no.~1, pp. 81--106, 1986.

\bibitem{Rokach2005}
L.~Rokach and O.~Maimon, ``{Top-down induction of decision trees classifiers -
  A survey},'' \emph{IEEE Transactions on Systems, Man and Cybernetics Part C:
  Applications and Reviews}, vol.~35, no.~4, pp. 476--487, 2005.

\bibitem{Quinlanpersonal}
\BIBentryALTinterwordspacing
{Quinlan, Ross}. {Ross Quinlan's personal homepage}. Accessed: 2018-06-03.
  [Online]. Available: \url{www.rulequest.com/Personal/}
\BIBentrySTDinterwordspacing

\bibitem{hoeting1999bayesian}
J.~A. Hoeting, D.~Madigan, A.~E. Raftery, and C.~T. Volinsky, ``Bayesian model
  averaging: a tutorial,'' \emph{Statistical science}, pp. 382--401, 1999.

\bibitem{d2016two}
J.~D'Hooge, D.~Barbosa, H.~Gao, P.~Claus, D.~Prater, J.~Hamilton, P.~Lysyansky,
  Y.~Abe, Y.~Ito, H.~Houle \emph{et~al.}, ``Two-dimensional speckle tracking
  echocardiography: standardization efforts based on synthetic ultrasound
  data,'' \emph{Eur Heart J Cardiovasc Imaging}, vol.~17, no.~6, pp. 693--701,
  2016.

\bibitem{Kearns1997}
M.~Kearns, Y.~Mansour, A.~Y. Ng, and D.~Ron, ``{An experimental and theoretical
  comparison of model selection methods},'' \emph{Machine Learning}, vol.~50,
  pp. 7--50, 1997.

\bibitem{Niblett1986}
T.~Niblett and I.~Bratko, ``Learning decision rules in noisy domains,'' in
  \emph{Proceedings of Expert Systems '86, The 6Th Annual Technical Conference
  on Research and development in expert systems III}.\hskip 1em plus 0.5em
  minus 0.4em\relax Cambridge University Press, 1986, pp. 25--34.

\bibitem{Lichman2013}
\BIBentryALTinterwordspacing
M.~Lichman, ``{UCI Machine Learning Repository},'' 2013. [Online]. Available:
  \url{http://archive.ics.uci.edu/ml}
\BIBentrySTDinterwordspacing

\bibitem{Alcala-Fdez2011}
J.~Alcal{\'{a}}-Fdez, A.~Fern{\'{a}}ndez, J.~Luengo, J.~Derrac,
  S.~Garc{\'{i}}a, L.~S{\'{a}}nchez, and F.~Herrera, ``{KEEL data-mining
  software tool: Data set repository, integration of algorithms and
  experimental analysis framework},'' \emph{Journal of Multiple-Valued Logic
  and Soft Computing}, vol.~17, no. 2-3, pp. 255--287, 2011.

\bibitem{Guyon2003}
\BIBentryALTinterwordspacing
I.~Guyon, ``Design of experiments for the nips 2003 variable selection
  benchmark,'' 2003. [Online]. Available:
  \url{clopinet.com/isabelle/Projects/NIPS2003}
\BIBentrySTDinterwordspacing

\bibitem{scikit-learn}
F.~Pedregosa, G.~Varoquaux, A.~Gramfort, V.~Michel, B.~Thirion, O.~Grisel,
  M.~Blondel, P.~Prettenhofer, R.~Weiss, V.~Dubourg, J.~Vanderplas, A.~Passos,
  D.~Cournapeau, M.~Brucher, M.~Perrot, and E.~Duchesnay, ``Scikit-learn:
  Machine learning in {P}ython,'' \emph{Journal of Machine Learning Research},
  vol.~12, pp. 2825--2830, 2011.

\bibitem{jensen1999effects}
D.~Jensen and T.~Oates, ``The effects of training set size on decision tree
  complexity,'' in \emph{Proceedings of the 14th International Conference on
  Machine Learning}, 1999, pp. 254--262.

\bibitem{kaggle2015}
\BIBentryALTinterwordspacing
{National Heart, Lung, and Blood Institute}, ``{Data Science Bowl Cardiac
  Challenge Data},'' 2015. [Online]. Available:
  \url{www.kaggle.com/c/second-annual-data-science-bowl}
\BIBentrySTDinterwordspacing

\bibitem{Ponikowski2016}
P.~Ponikowski \emph{et~al.}, ``{2016 ESC Guidelines for the diagnosis and
  treatment of acute and chronic heart failure: The Task Force for the
  diagnosis and treatment of acute and chronic heart failure of the European
  Society of Cardiology~(ESC) Developed with the special contribution of the
  Heart Failure Association~(HFA) of the~ESC},'' \emph{{European heart
  journal}}, vol.~37, no.~27, pp. 2129--2200, 2016.

\bibitem{Demsar2006}
J.~Demsar, ``{Statistical Comparison of Classifiers over Multiple Data Sets},''
  \emph{Journal of Machine Learning Research}, vol.~7, no.~7, pp. 1--30, 2006.

\bibitem{Wilcoxon1945}
F.~Wilcoxon, ``{Individual Comparisons by Ranking Methods},'' \emph{Biometrics
  Bulletin}, vol.~1, no.~6, pp. 80--83, 1945.

\bibitem{Clopper1934}
C.~Clopper and E.~Pearson, ``The use of confidence or fiducial limits
  illustrated in the case of the binomial,'' \emph{Biometrika}, vol.~26, no.~4,
  p. 404, 1934.

\end{thebibliography}

\clearpage

\appendices

\section{Parameter tuning}

\renewcommand{\thefigure}{A.\arabic{figure}} \setcounter{figure}{0}

Figures~\ref{fig:params-train-noise} and \ref{fig:params-test-noise} display the average value of the parameters that control the uncertainty distribution for \ac{SS}, \ac{STP}, \ac{SE} and \ac{UDT}, selected through~\ac{CV}.

\section{Pruning algorithm in C4.5} \label{appendixA}

\renewcommand{\thefigure}{B.\arabic{figure}} \setcounter{figure}{0}

C4.5 employs the following pruning algorithm. Consider a leaf with majority class $y$, where an instance misclassification follows a Bernoulli distribution with probability $p(\neg y)$. We now show how to obtain an estimate $\hat{p}$ of $p(\neg y)$ from the training data. During training, the leaf sees $N$ instances, $N(y)$ belong to class~$y$. Let us assume that the number of training errors $N(\neg y)$ follows a binomial distribution, $N(\neg y) \sim B\big(N, p(\neg y)\big)$. One approach to estimating $p(\neg y)$ is to consider the upper limit of the confidence interval $[p_L,p_U]$ of this binomial under a specified confidence level, $\hat{p} = p_U$. In C4.5, a two-tailed confidence level is set by specifying the \textit{confidence factor} $c$, such that the probability that $p(\neg y) > \hat{p}$ is smaller than $\tfrac{c}{2}$. To estimate the binomial confidence limits, the Clopper-Pearson method was used~\cite{Clopper1934}.

The number of errors is finally estimated by multiplying the number of instances by the probability estimate, $\hat{N}(\neg y) = N \times p_U$. This is computed for three scenarios: 1) maintaining the node 2) replacing the node by a leaf and 3) replacing the node by the subbranch with smallest predicted error. The smallest-error scenario is chosen. If $c$ is small, $\hat{p}$ will be higher for leaves with less instances, and the tree will be more aggressively pruned. If $c$ is high, $\hat{p}$ will be smaller for nodes with few instances compared to the parent. The tree will be less pruned.


\section{Changing the pruning confidence factor} \label{appendixB}

\renewcommand{\thefigure}{C.\arabic{figure}} \setcounter{figure}{0}

To complement the results, we investigate if the reduction in the number of leaves with maintained accuracy could have been achieved using C4.5 with a distinct pruning confidence factor $c$, in order to achieve the same regularizing effect. We evaluated the models learned using C4.5 and \ac{SS} for $c \in [0,1]$ with noise levels $n_{train}=0$ and $n_{train}=0.1$. The results can be seen in Figures \ref{fig:cf-n-0} and \ref{fig:cf-n-01} for 5 of the datasets. Both in the case with no-added noise and with $n=0.1$, the \ac{SS} approach led to smaller trees for the lower range of $c$. The differences in accuracy between the two methods were small. This indicates that for the lower range of $c$, \ac{SS} was able to make more assertive estimates of the generalization error compared to C4.5.


\section{Absolute results} \label{abs-res}

This section of the Appendix shows the absolute value of the accuracy and number of leaves for each, obtained with train and test noise levels~0.0 and 0.2.

\renewcommand{\thefigure}{A.\arabic{figure}} \setcounter{figure}{0}

\begin{figure}[h]
	\captionsetup[subfigure]{justification=centering} \centering
	\includegraphics[width=.24\textwidth]{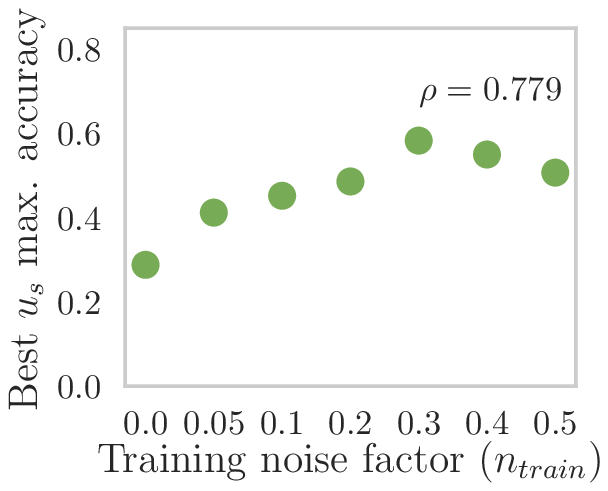}
	\includegraphics[width=.24\textwidth]{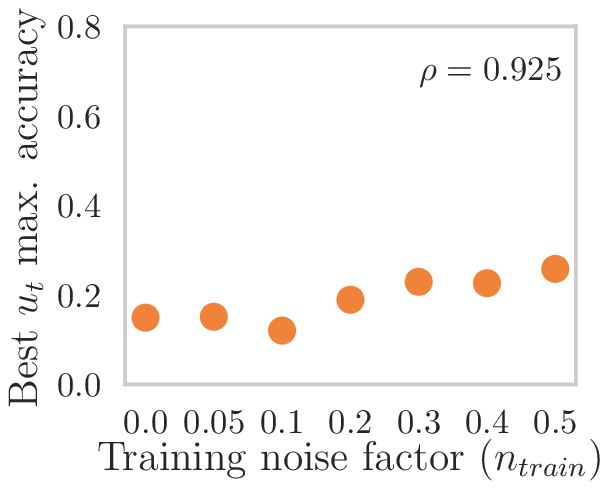}
	\includegraphics[width=.24\textwidth]{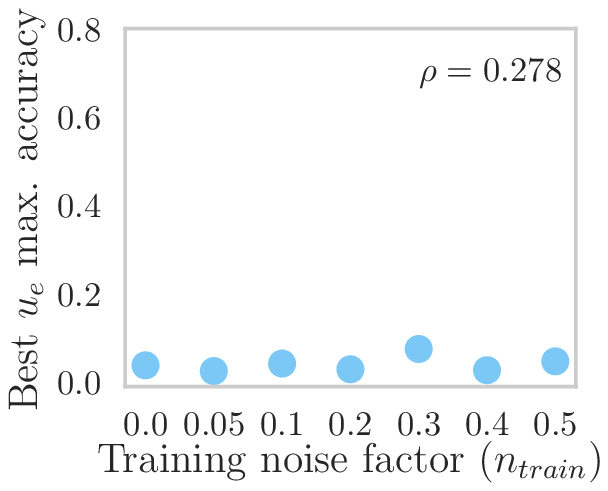}
	\includegraphics[width=.24\textwidth]{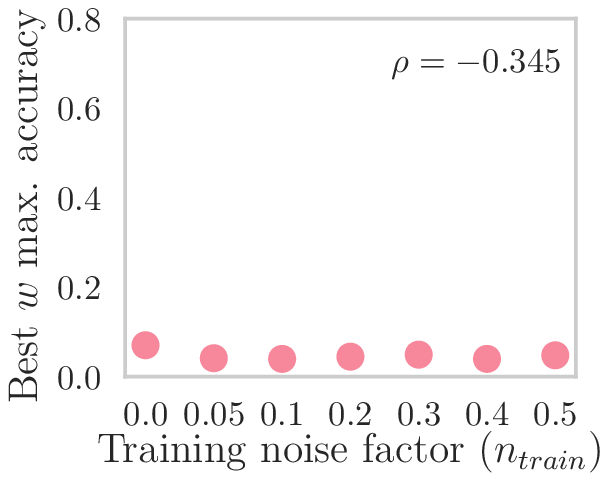}
	\caption{Average uncertainty factors ($u_s$, $u_t$, $u_e$) and window factor ($w$) selected to maximize \acf{CV} accuracy in the presence of noise added to the \ac{CV} training folds.} \label{fig:params-train-noise}
\end{figure}
\begin{figure}[t]
	\captionsetup[subfigure]{justification=centering} \centering
	
	\includegraphics[width=.24\textwidth]{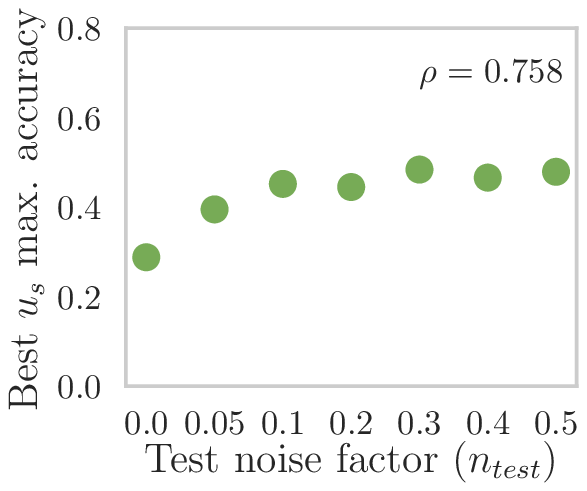}
	\includegraphics[width=.24\textwidth]{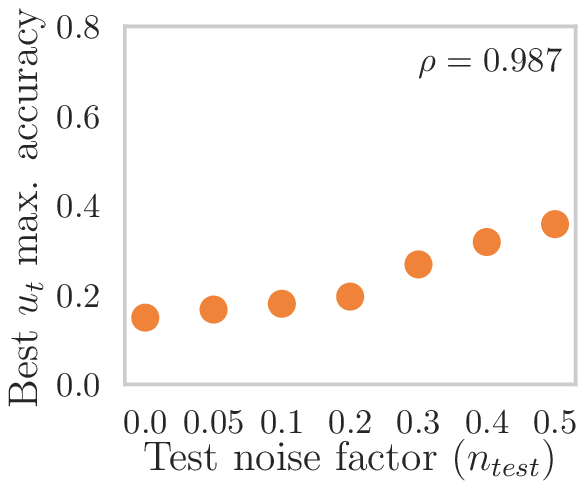}
	\includegraphics[width=.24\textwidth]{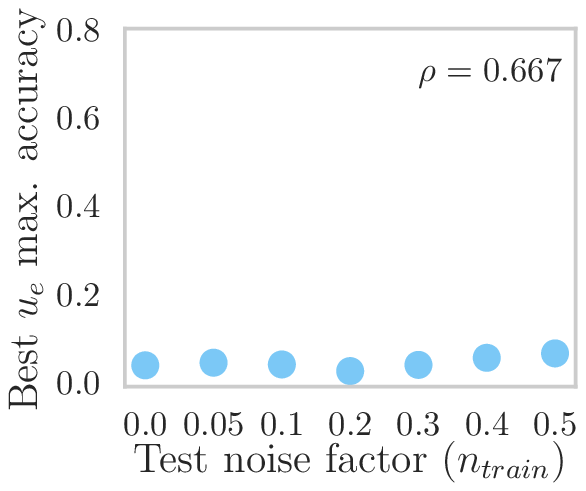}
	\includegraphics[width=.24\textwidth]{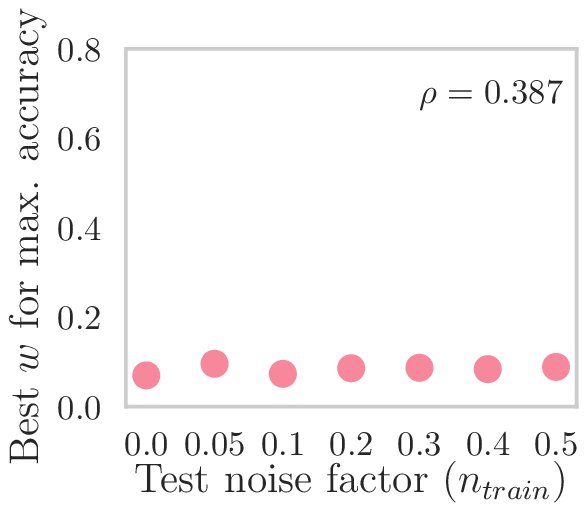}
	\caption{Average uncertainty factors ($u_s$, $u_t$, $u_e$) and window factor ($w$) selected to maximize \acf{CV} accuracy in the presence of noise added the \ac{CV} validation folds.} \label{fig:params-test-noise}
	\vspace{6cm}
\end{figure}

\vfill

\clearpage

\renewcommand{\thefigure}{C.\arabic{figure}} \setcounter{figure}{0}

\begin{figure}
	\captionsetup[subfigure]{justification=centering} \centering
	\begin{subfigure}{.5\textwidth}\centering
		\includegraphics[width=\textwidth]{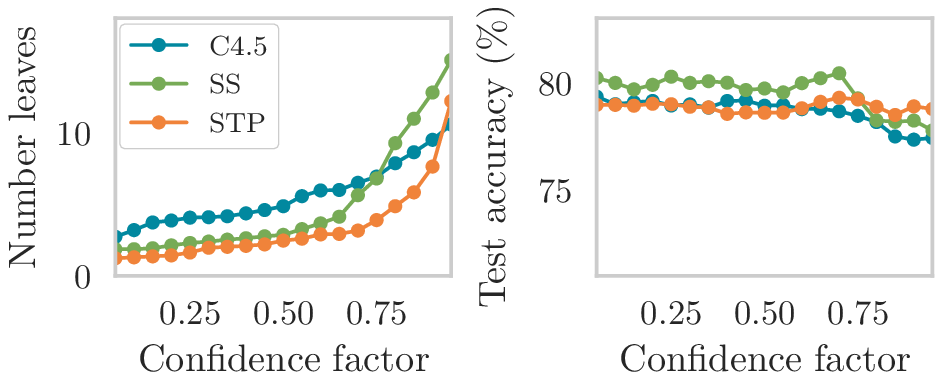}
		\caption{ \scriptsize Hepatitis dataset \\~\\}\vspace{-3mm}
	\end{subfigure}
	
	\begin{subfigure}{.5\textwidth}\centering
		\includegraphics[width=\textwidth]{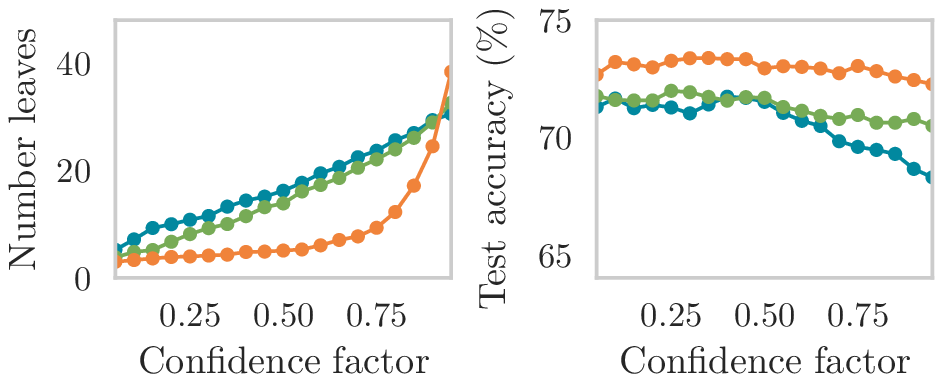}
		\caption{ \scriptsize Heart disease dataset \\~\\}\vspace{-3mm}
	\end{subfigure}
	
	\begin{subfigure}{.5\textwidth}\centering
		\includegraphics[width=\textwidth]{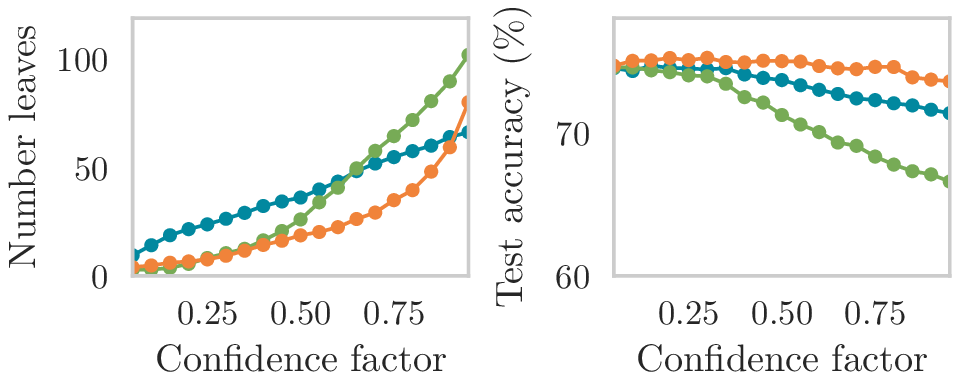} \caption{
			\scriptsize Pima Indians diabetes dataset \\~\\}\vspace{-3mm}
	\end{subfigure}
	
	\begin{subfigure}{.5\textwidth}\centering
		\includegraphics[width=\textwidth]{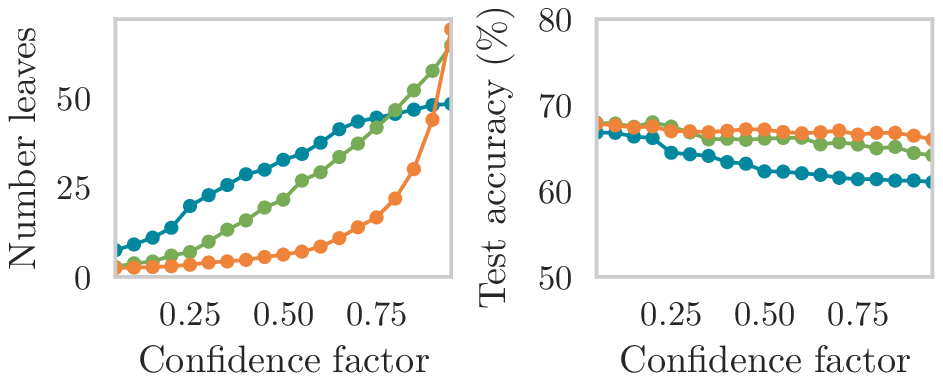} \caption{
			\scriptsize South African heart disease dataset \\~\\}\vspace{-3mm}
	\end{subfigure}
	
	\begin{subfigure}{.5\textwidth}\centering
		\includegraphics[width=\textwidth]{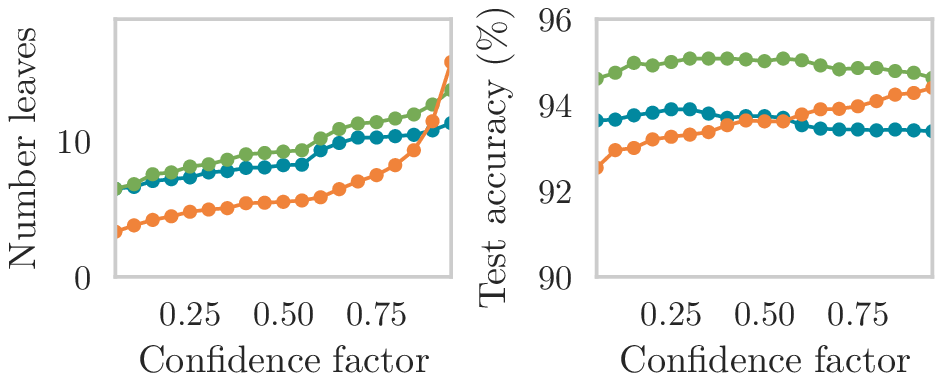}
		\caption{ \scriptsize Breast cancer dataset \\~\\}\vspace{-3mm}
	\end{subfigure}
	\caption{ \footnotesize Effect of varying the pruning confidence factor using C4.5, \acf{SS} or \acf{STP} without added noise and optimal $u_s$ and $u_t$, selected through \acf{CV}.}
	\label{fig:cf-n-0}
\end{figure}

\begin{figure}
	\captionsetup[subfigure]{justification=centering} \centering
	\begin{subfigure}{.5\textwidth}\centering
		\includegraphics[width=\textwidth]{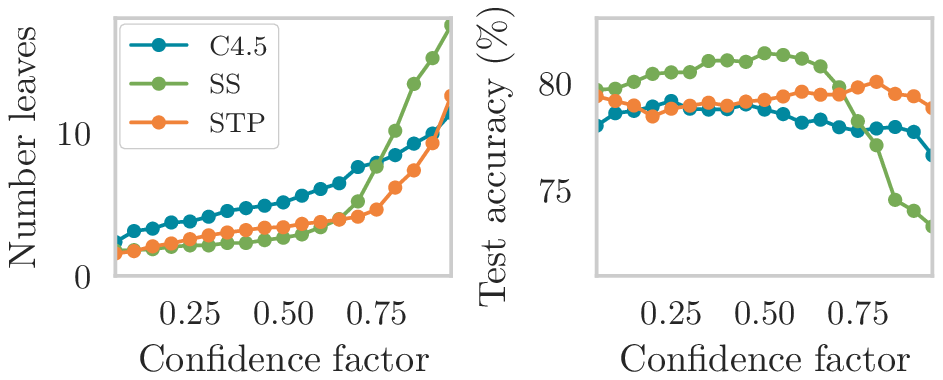}
		\caption{ \scriptsize Hepatitis dataset \\~\\}\vspace{-3mm}
	\end{subfigure}
	
	\begin{subfigure}{.5\textwidth}\centering
		\includegraphics[width=\textwidth]{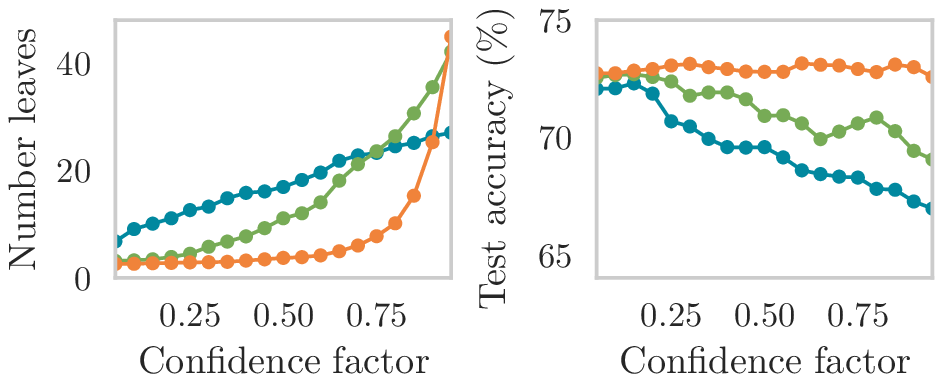}
		\caption{ \scriptsize Heart disease dataset \\~\\}\vspace{-3mm}
	\end{subfigure}
	
	\begin{subfigure}{.5\textwidth}\centering
		\includegraphics[width=\textwidth]{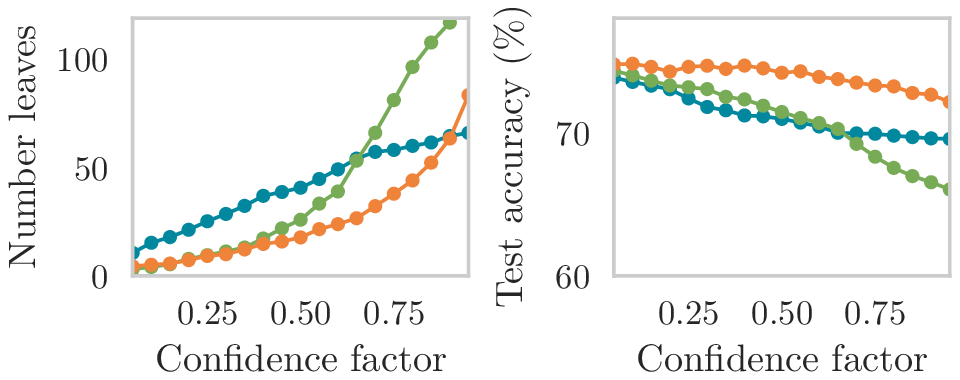} \caption{
			\scriptsize Pima Indians diabetes dataset \\~\\}\vspace{-3mm}
	\end{subfigure}
	
	\begin{subfigure}{.5\textwidth}\centering
		\includegraphics[width=\textwidth]{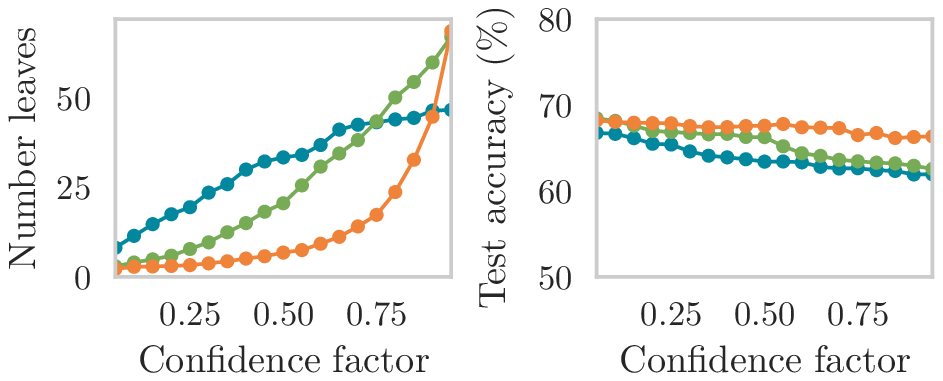} \caption{
			\scriptsize South African heart disease dataset \\~\\}\vspace{-3mm}
	\end{subfigure}
	
	\begin{subfigure}{.5\textwidth}\centering
		\includegraphics[width=\textwidth]{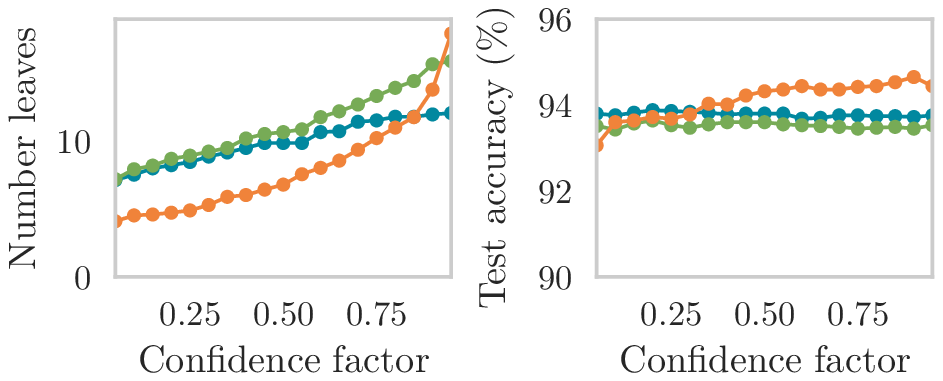}
		\caption{ \scriptsize Breast cancer dataset \\~\\}\vspace{-3mm}
	\end{subfigure}
	\caption{ \footnotesize Effect of varying the pruning confidence factor using C4.5, \acf{SS} or \acf{STP} with $n_{train} = 0.1$ and optimal $u_s$ and $u_t$, selected through \acf{CV}.}
	\label{fig:cf-n-01}
\end{figure}

	\begin{table*}[h!]
	\centering
	\setlength{\tabcolsep}{4pt}
	\renewcommand{\arraystretch}{1.3}
	\caption{ \footnotesize Average absolute number of leaves per dataset in Experiment 1, with $n_{train}=0.00$ and $n_{train}=0.20$.}
	\scriptsize
	\begin{tabular}{l|rrrrrr|rrrrrr}
		\midrule
		\multirow{2}{*}{Dataset} 	& \multicolumn{6}{c}{$n_{train}=0.00$} & \multicolumn{6}{c}{$n_{train}=0.20$} \\
		& C4.5 & SS & STP & SE & PLT & UDT & C4.5 & SS & STP & SE & PLT & UDT \\
		\midrule \midrule
		
		Hepatitis				&	5.0		&	3.0		&	2.5		&	5.0		&	5.0		&	5.0		&	5.2		&	2.5		&	1.5		&	5.2		&	5.2		&	5.3		\\
		Heart disease			&	15.0	&	13.1	&	4.9		&	15.0	&	15.0	&	15.8	&	16.4	&	8.5		&	3.6		&	16.4	&	16.4	&	15.8	\\
		Pima Indians diabetes	&	15.2	&	2.9		&	5.0		&	15.2	&	15.2	&	19.1	&	17.5	&	13.5	&	4.0		&	17.5	&	17.5	&	20.2	\\
		South African heart		&	15.0	&	6.1		&	3.0		&	15.0	&	15.0	&	15.3	&	18.0	&	6.9		&	3.8		&	18.0	&	18.0	&	19.2	\\
		Breast cancer Wisconsin	&	10.0	&	12.8	&	11.6	&	10.0	&	10.0	&	12.2	&	13.6	&	14.9	&	17.4	&	13.6	&	13.6	&	14.4	\\
		Dermatology				&	10.1	&	11.9	&	21.7	&	10.1	&	10.1	&	10.2	&	10.9	&	10.6	&	11.4	&	10.9	&	10.9	&	10.2	\\
		Haberman				&	15.1	&	10.0	&	2.0		&	15.1	&	15.1	&	19.9	&	18.4	&	10.3	&	5.6		&	18.4	&	18.4	&	20.2	\\
		Indian liver patient	&	15.5	&	1.4		&	1.0		&	15.5	&	15.5	&	15.5	&	17.7	&	1.5		&	1.0		&	17.7	&	17.7	&	19.2	\\
		BUPA liver disorders	&	15.2	&	10.7	&	9.1		&	15.2	&	15.2	&	15.2	&	16.9	&	10.1	&	6.4		&	16.9	&	16.9	&	18.4	\\
		Vertebral column (2c)	&	15.1	&	15.1	&	11.8	&	15.1	&	15.1	&	15.1	&	16.5	&	17.8	&	13.6	&	16.5	&	16.5	&	17.8	\\
		Vertebral column (3c)	&	15.1	&	12.4	&	11.4	&	15.1	&	15.1	&	15.0	&	17.4	&	16.1	&	11.1	&	17.4	&	17.4	&	16.1	\\
		Thyroid gland			&	5.0		&	5.4		&	5.2		&	5.0		&	5.0		&	5.0		&	5.6		&	5.1		&	4.7		&	5.6		&	5.6		&	5.3		\\
		Oxford PD				&	5.0		&	5.3		&	2.3		&	5.0		&	5.0		&	5.4		&	5.7		&	4.9		&	2.9		&	5.7		&	5.7		&	7.4		\\
		SPECTF					&	15.0	&	20.0	&	4.3		&	15.0	&	15.0	&	14.9	&	15.5	&	23.7	&	4.1		&	15.5	&	15.5	&	15.4	\\
		Thoracic surgery		&	15.5	&	3.6		&	1.0		&	15.5	&	15.5	&	23.4	&	18.5	&	3.1		&	1.0		&	18.5	&	18.5	&	21.9	\\
		Synthetic 1				&	15.2	&	15.2	&	15.2	&	15.2	&	15.2	&	21.9	&	20.4	&	7.3		&	6.5		&	20.4	&	20.4	&	22.9	\\
		Synthetic 2				&	15.0	&	15.0	&	15.0	&	15.0	&	15.0	&	15.3	&	25.7	&	18.5	&	5.1		&	25.7	&	25.7	&	25.7	\\
		Synthetic 3				&	15.6	&	12.2	&	7.8		&	15.6	&	15.6	&	17.0	&	20.6	&	12.3	&	8.0		&	20.6	&	20.6	&	19.6	\\
		Synthetic 4				&	15.2	&	12.7	&	6.1		&	15.2	&	15.2	&	16.5	&	20.0	&	12.8	&	11.4	&	20.0	&	20.0	&	19.6	\\
		Synthetic 5				&	15.2	&	15.2	&	15.2	&	15.2	&	15.2	&	16.4	&	17.6	&	9.4		&	9.1		&	17.6	&	17.6	&	18.3	\\
		
		\midrule
		Average difference to C4.5 & 		&	-2.9	&	-5.4	&	0.0		&	0.0		&	1.5		&			&	-5.4	&	-9.3	&	0.0		&	0.0		&	0.7		\\
		\midrule
	\end{tabular} 
	\label{table:abs-lvs}
\end{table*}	

\begin{table*}[h!]
	\centering
	\setlength{\tabcolsep}{4pt}
	\renewcommand{\arraystretch}{1.3}
	\caption{ \footnotesize Average absolute test accuracy (\%) per dataset in Experiment 1, with $n_{train}=0.00$ and $n_{train}=0.20$.}
	\scriptsize
	\begin{tabular}{l|rrrrrr|rrrrrr}
		\midrule
		\multirow{2}{*}{Dataset} 	& \multicolumn{6}{c|}{$n_{train}=0.00$} & \multicolumn{6}{c}{$n_{train}=0.20$} \\
		& C4.5 & SS & STP & SE & PLT & UDT & C4.5 & SS & STP & SE & PLT & UDT \\
		\midrule \midrule
		Hepatitis					&	79.0	&	79.6	&	78.6	&	78.4	&	78.3	&	79.0	& 78.9	& 80.8	& 79.7	& 79.2	& 79.2	& 79.4	\\
		Heart disease				&	71.6	&	71.6	&	73.3	&	68.4	&	68.4	&	71.4	& 69.9	& 72.0	& 72.6	& 71.5	& 69.8	& 71.4	\\
		Pima Indians diabetes		&	74.3	&	74.5	&	75.0	&	74.5	&	74.3	&	74.3	& 71.8	& 72.7	& 73.3	& 71.6	& 71.4	& 71.8	\\
		South African heart			&	65.5	&	67.9	&	67.5	&	65.9	&	65.5	&	66.7	& 65.4	& 67.5	& 67.7	& 65.0	& 65.4	& 66.1	\\
		Breast cancer Wisconsin		&	93.2	&	94.1	&	94.0	&	93.3	&	93.0	&	94.6	& 92.3	& 92.4	& 93.3	& 93.2	& 92.4	& 93.0	\\
		Dermatology					&	95.7	&	95.7	&	95.6	&	86.9	&	83.4	&	95.7	& 94.8	& 95.4	& 95.2	& 94.3	& 92.7	& 95.4	\\
		Haberman					&	71.4	&	73.6	&	73.0	&	71.9	&	71.8	&	69.5	& 69.9	& 73.3	& 73.9	& 70.3	& 70.2	& 71.6	\\
		Indian liver patient		&	68.2	&	70.9	&	71.0	&	68.1	&	68.1	&	68.2	& 67.5	& 70.6	& 71.0	& 67.7	& 67.7	& 67.9	\\
		BUPA liver disorders		&	66.1	&	66.9	&	64.6	&	65.2	&	64.3	&	66.1	& 60.6	& 64.1	& 62.1	& 60.1	& 60.6	& 60.4	\\
		Vertebral column (2c)		&	79.1	&	79.1	&	79.6	&	79.0	&	79.0	&	79.1	& 78.2	& 77.0	& 77.5	& 77.7	& 77.2	& 77.5	\\
		Vertebral column (3c)		&	81.4	&	80.5	&	82.4	&	81.2	&	80.4	&	81.0	& 76.2	& 78.8	& 79.3	& 77.9	& 76.0	& 77.8	\\
		Thyroid gland				&	92.8	&	92.5	&	92.8	&	92.5	&	91.4	&	92.8	& 91.7	& 90.4	& 90.7	& 90.7	& 90.0	& 90.8	\\
		Oxford PD					&	83.6	&	84.2	&	82.4	&	83.0	&	83.4	&	84.8	& 79.7	& 80.5	& 81.7	& 81.2	& 79.8	& 80.3	\\
		SPECTF						&	74.0	&	75.8	&	79.0	&	75.0	&	74.9	&	74.1	& 70.3	& 74.6	& 79.0	& 71.7	& 71.0	& 71.6	\\
		Thoracic surgery			&	80.1	&	84.1	&	85.1	&	79.4	&	80.1	&	82.3	& 80.3	& 84.6	& 85.1	& 79.0	& 80.5	& 83.6	\\
		Synthetic 1					&	72.6	&	72.6	&	72.6	&	63.7	&	71.8	&	72.2	& 52.6	& 51.7	& 52.0	& 52.2	& 52.7	& 52.3	\\
		Synthetic 2					&	77.8	&	77.8	&	77.8	&	68.4	&	77.2	&	80.1	& 56.7	& 55.1	& 54.8	& 53.7	& 55.5	& 53.8	\\
		Synthetic 3					&	69.1	&	69.2	&	69.0	&	67.8	&	68.2	&	68.6	& 52.1	& 55.7	& 56.9	& 54.7	& 52.1	& 55.4	\\
		Synthetic 4					&	54.6	&	54.6	&	55.2	&	49.9	&	54.1	&	55.5	& 36.9	& 37.2	& 35.4	& 33.6	& 35.4	& 33.4	\\
		Synthetic 5					&	57.0	&	57.0	&	57.0	&	55.4	&	55.5	&	57.1	& 40.5	& 44.6	& 43.0	& 39.2	& 40.1	& 40.4	\\
		\midrule
		Avg. difference to C4.5		&  		&	0.8		&	0.9		&	-2.0	&	-1.2	&	0.3		& 		& 1.6	& 1.9	& -0.1	& -0.3	& 0.4	\\			
		\midrule
	\end{tabular}
	\label{table:abs-acc-train-noise}
\end{table*}

\begin{table*}[h!]
	\centering
	\setlength{\tabcolsep}{4pt}
	\renewcommand{\arraystretch}{1.3}
	\caption{ \footnotesize Average absolute test accuracy (\%) per dataset in Experiment 2, with $n_{test}=0.00$ and $n_{test}=0.20$.}
	\scriptsize
	\begin{tabular}{l|rrrrrr|rrrrrr}
		\midrule
		\multirow{2}{*}{Dataset} 	& \multicolumn{6}{c}{$n_{test}=0.00$} & \multicolumn{6}{|c}{$n_{test}=0.20$} \\
		& C4.5 & SSS & STP & SE & PLT & UDT & C4.5 & SSS & STP & SE & PLT & UDT \\
		\midrule \midrule
		Hepatitis & 79.0 & 79.6 & 78.6 & 78.4 & 78.3 & 79.0 & 75.2 & 79.3 & 78.9 & 75.2 & 75.3 & 75.2 \\
		Heart disease & 71.6 & 71.6 & 73.3 & 68.4 & 68.4 & 71.4 & 66.5 & 67.8 & 71.7 & 66.0 & 65.8 & 65.6 \\
		Pima Indians diabetes & 74.3 & 74.5 & 75.0 & 74.5 & 74.3 & 74.3 & 69.3 & 69.3 & 71.0 & 69.6 & 69.8 & 69.3 \\
		South African heart & 65.5 & 67.9 & 67.5 & 65.9 & 65.6 & 66.7 & 66.1 & 67.2 & 67.8 & 66.2 & 65.9 & 66.1 \\
		Breast cancer Wisconsin & 93.2 & 94.2 & 94.0 & 93.3 & 93.0 & 94.6 & 86.5 & 90.9 & 90.2 & 86.4 & 86.1 & 87.3 \\
		Dermatology & 95.7 & 95.7 & 95.6 & 86.9 & 83.4 & 95.7 & 87.0 & 89.6 & 89.5 & 87.0 & 79.9 & 95.5 \\
		Haberman & 71.4 & 73.6 & 73.0 & 71.9 & 71.8 & 69.5 & 71.3 & 71.0 & 74.2 & 71.0 & 71.1 & 71.3 \\
		Indian liver patient & 68.2 & 70.9 & 71.0 & 68.1 & 68.1 & 68.2 & 68.5 & 70.9 & 71.0 & 68.4 & 68.3 & 68.5 \\
		BUPA liver disorders & 66.1 & 66.9 & 64.7 & 65.2 & 64.3 & 66.1 & 61.1 & 61.4 & 63.0 & 61.0 & 61.0 & 61.1 \\
		Vertebral column (2c) & 79.1 & 79.1 & 79.6 & 79.0 & 79.0 & 79.1 & 75.8 & 75.8 & 77.3 & 75.8 & 75.3 & 75.8 \\
		Vertebral column (3c) & 81.4 & 80.5 & 82.4 & 81.2 & 80.4 & 81.0 & 75.5 & 76.6 & 77.5 & 75.6 & 74.3 & 75.6 \\
		Thyroid gland & 92.8 & 92.5 & 92.8 & 92.5 & 91.4 & 92.8 & 83.2 & 88.4 & 85.3 & 83.0 & 82.6 & 85.3 \\
		Oxford PD & 83.6 & 84.2 & 82.4 & 83.0 & 83.4 & 84.8 & 77.6 & 78.6 & 77.8 & 77.5 & 77.3 & 76.5 \\
		SPECTF & 74.0 & 75.8 & 79.0 & 75.0 & 74.9 & 74.1 & 72.4 & 75.2 & 79.0 & 72.4 & 72.1 & 72.9 \\
		Thoracic surgery & 80.1 & 84.1 & 85.1 & 79.4 & 80.2 & 82.3 & 78.7 & 78.8 & 85.1 & 78.3 & 78.6 & 78.9 \\
		Synthetic 1 & 72.6 & 72.6 & 72.6 & 63.7 & 71.8 & 72.2 & 51.6 & 52.2 & 52.8 & 51.4 & 51.5 & 52.4 \\
		Synthetic 2 & 77.8 & 77.8 & 77.8 & 59.6 & 77.2 & 80.1 & 51.4 & 54.0 & 53.2 & 50.9 & 51.5 & 51.4 \\
		Synthetic 3 & 69.1 & 69.2 & 69.0 & 67.8 & 68.2 & 68.6 & 53.9 & 55.4 & 55.7 & 54.0 & 53.7 & 53.4 \\
		Synthetic 4 & 54.6 & 54.7 & 55.3 & 49.9 & 54.1 & 55.5 & 35.4 & 39.4 & 36.3 & 35.4 & 35.3 & 35.9 \\
		Synthetic 5 & 57.0 & 57.0 & 57.0 & 55.4 & 55.6 & 57.1 & 39.5 & 41.2 & 39.6 & 39.7 & 39.5 & 38.0 \\
		\midrule
		Avg. difference to C4.5		&  		&	0.8		&	0.9		&	-2.0	&	-1.2	&	0.3		& 		& 1.8	&	2.5	& -0.1	& -0.6	& 0.5	\\
		\midrule
	\end{tabular}
	\label{table:abs-acc-test-noise}
\end{table*}

\begin{table*}[h!]
	\centering
	\setlength{\tabcolsep}{2.2pt}
	\renewcommand{\arraystretch}{1.3}
	\caption{ \footnotesize Average absolute learning and evaluation time (s) per dataset in Experiment 1, with $n_{train}=0.00$ and $n_{train}=0.20$.}
	\scriptsize
	\begin{tabular}{l|rrrrrr|rrrrrr}
		\midrule
		\multirow{2}{*}{Dataset} 	& \multicolumn{6}{|c|}{$n_{train}=0.00$} & \multicolumn{6}{|c}{$n_{train}=0.20$} \\
		& C4.5 & SSS & STP & SE & PLT & UDT & C4.5 & SSS & STP & SE & PLT & UDT \\ \midrule \midrule
		
		Hepatitis & 0.03 & 1.60 & 0.13 & 0.02 & 0.07 & 0.02 & 0.06 & 3.45 & 0.20 & 0.05 & 0.11 & 0.03 \\
		Heart disease & 0.04 & 1.24 & 0.33 & 0.04 & 0.11 & 4.80 & 0.22 & 2.82 & 1.31 & 0.11 & 0.25 & 4.80 \\
		Pima Indians diabetes & 0.29 & 6.98 & 1.26 & 0.16 & 0.37 & 26.19 & 0.61 & 9.51 & 4.71 & 0.35 & 0.83 & 132.99 \\
		South African heart & 0.31 & 4.23 & 1.30 & 0.13 & 0.31 & 27.69 & 0.40 & 4.44 & 1.75 & 0.18 & 0.43 & 56.98 \\
		Breast cancer Wisconsin & 0.83 & 7.93 & 10.63 & 0.44 & 1.06 & 287.86 & 0.94 & 10.10 & 17.47 & 0.51 & 1.15 & 227.31 \\
		Dermatology & 0.10 & 5.57 & 1.18 & 0.06 & 0.19 & 3.31 & 1.06 & 10.22 & 0.98 & 0.62 & 1.51 & 157.46 \\
		Haberman & 0.03 & 0.51 & 0.44 & 0.02 & 0.04 & 1.34 & 0.07 & 1.57 & 0.28 & 0.05 & 0.09 & 12.62 \\
		Indian liver patient & 0.19 & 6.56 & 1.30 & 0.17 & 0.23 & 0.14 & 0.37 & 10.10 & 3.07 & 0.30 & 0.52 & 0.31 \\
		BUPA liver disorders & 0.09 & 1.75 & 0.09 & 0.04 & 0.08 & 0.05 & 0.15 & 1.97 & 0.69 & 0.10 & 0.19 & 21.91 \\
		Vertebral column (2c) & 0.19 & 0.19 & 0.34 & 0.09 & 0.21 & 0.09 & 0.18 & 2.37 & 0.36 & 0.10 & 0.25 & 33.51 \\
		Vertebral column (3c) & 0.10 & 1.26 & 0.13 & 0.06 & 0.13 & 17.18 & 0.16 & 1.67 & 0.36 & 0.07 & 0.15 & 0.08 \\
		Thyroid gland & 0.02 & 1.13 & 0.02 & 0.01 & 0.02 & 0.01 & 0.06 & 0.79 & 0.06 & 0.03 & 0.06 & 7.40 \\
		Oxford PD & 0.11 & 1.45 & 0.80 & 0.07 & 0.16 & 26.18 & 0.16 & 1.94 & 0.66 & 0.09 & 0.21 & 33.04 \\
		SPECTF & 0.20 & 6.62 & 2.19 & 0.12 & 0.25 & 9.70 & 0.52 & 5.61 & 6.38 & 0.34 & 0.76 & 0.34 \\
		Thoracic surgery & 0.11 & 2.61 & 0.45 & 0.05 & 0.12 & 6.10 & 0.20 & 5.97 & 0.87 & 0.08 & 0.28 & 29.22 \\
		Synthetic 1 & 0.64 & 0.64 & 0.34 & 0.34 & 1.00 & 140.76 & 1.28 & 10.52 & 1.52 & 0.52 & 1.65 & 137.90 \\
		Synthetic 2 & 0.47 & 0.47 & 0.23 & 0.23 & 0.80 & 98.04 & 0.91 & 9.36 & 3.79 & 0.40 & 1.20 & 89.28 \\
		Synthetic 3 & 0.39 & 3.76 & 1.37 & 0.21 & 0.75 & 71.05 & 0.53 & 4.91 & 1.31 & 0.30 & 0.91 & 0.35 \\
		Synthetic 4 & 0.37 & 3.40 & 1.42 & 0.18 & 0.81 & 53.46 & 0.53 & 4.68 & 0.68 & 0.23 & 0.85 & 0.29 \\
		Synthetic 5 & 0.41 & 0.41 & 0.20 & 0.19 & 0.71 & 42.43 & 0.61 & 5.88 & 0.72 & 0.26 & 1.03 & 43.77 \\ 
		\midrule
		Avg. difference to C4.5	& 	& 2.67	& 0.96	& -0.12	& 0.12	& 40.57	& 	&	4.94	&	1.91	&	-0.22	&	0.17	&	49.03	\\
		\midrule
	\end{tabular}
	\label{table:abs-time}
\end{table*}

\end{document}